\documentclass[journal]{journal}
\pdfoutput=1
\usepackage{multicol,lipsum}
\usepackage{booktabs}
\usepackage[normalem]{ulem}
\useunder{\uline}{\ul}{}

\ifCLASSINFOpdf
	\usepackage[pdftex]{graphicx}
 
\else
\fi
\usepackage[cmex10]{amsmath}

\usepackage{mathtools}

\usepackage{float}

\usepackage[export]{adjustbox}

\usepackage{fixltx2e}
\usepackage{ulem}
\usepackage{subfigure}
\usepackage{xcolor}
\usepackage{setspace}

\usepackage[colorlinks,
            linkcolor=blue,
            anchorcolor=blue,
            citecolor=blue]{hyperref}

\usepackage{nameref}

\begin{document}
%
\title{Newtonian Mechanics Based Transient Stability PART II: Individual Machine}
%
%
%

\author{{Songyan Wang,
        Jilai Yu,  
				Aoife Foley,
        Jingrui Zhang
        }
        
}
%
%

\markboth{Journal of \LaTeX\ Class Files,~Vol.~6, No.~1, January~2007}%
{Shell \MakeLowercase{\textit{et al.}}: Bare Demo of IEEEtran.cls for Journals}
%



\maketitle
\thispagestyle{empty}
\begin{abstract}

The paper analyzes the mechanisms of the individual-machine and also its advantages in TSA. Based on the critical-machine monitoring of the original system trajectory, it is clarified that the individual-machine strictly follows the machine paradigms. 
These strict followings of the paradigms bring the two advantages of the individual-machine method in TSA: (i) the individual-machine trajectory stability is characterized precisely, and (ii) the individual-machine trajectory variance is depicted clearly at IMPP. The two advantages are fully reflected in the precise definitions of individual-machine based transient stability concepts. 
In particular, the critical machine swing is clearly depicted through the IDSP or IDLP of the machine, the critical stability of the system is strictly defined as the critical stability of the most-severely disturbed machine, and the individual-machine potential energy surface is also precisely modeled through the IMPE of the machine. Simulation results show the effectiveness of the individual-machine in TSA.

\end{abstract}

%

\IEEEpeerreviewmaketitle

  \begin{tabular}{lllll}
    &            &               &                  &                                \\
  \multicolumn{5}{c}{\textbf{Nomenclature}}                                                \\
  KE     & \multicolumn{1}{c}{}  & \multicolumn{3}{l}{Kinetic energy}                      \\
  PE     & \multicolumn{1}{c}{}  & \multicolumn{3}{l}{Potential energy}                    \\
  COI    & \multicolumn{1}{c}{}  & \multicolumn{3}{l}{Center of inertia}                   \\
  DLP    &                       & \multicolumn{3}{l}{Dynamic liberation point}            \\
  DSP    &                       & \multicolumn{3}{l}{Dynamic stationary point}            \\
  EAC    &                       & \multicolumn{3}{l}{Equal area criterion}                \\
  MPP    &                       & \multicolumn{3}{l}{Maximum potential energy point}      \\
  NEC    &                       & \multicolumn{3}{l}{Newtonian energy conversion}         \\
  TSA    &                       & \multicolumn{3}{l}{Transient stability assessment}      \\
  TEF    &                       & \multicolumn{3}{l}{Transient energy function}           \\
  SMTE   &                       & \multicolumn{3}{l}{Superimposed-machine transient energy}   \\
  \end{tabular}

\section{Introduction}

%
%
%
%
\subsection{LITERATURE REVIEW}
\raggedbottom
\IEEEPARstart{I}{ndividual} machine is physically real and it is also the fundamental component of a multi-machine power system. For the power system that suffers severe disturbance, the trajectory variance of the entire system is originally depicted through the trajectory variance of each physically real individual machine.
The individual-machine thinking was originally formed by Fouad and Stanton \cite{1}, \cite{2}. In the two paper series, Fouad and Stanton stated that ``the instability of a multi-machine system is determined by the motion of some unstable critical machines if more than one machine tends to lose synchronism”. Stimulated by this individual-machine perspective, Vittal and Fouad stated that the instability of the system was decided by the individual-machine transient energy, and in this way the transient stability of the entire multi-machine system was depicted in a genuine individual-machine manner \cite{3}, \cite{4}.
The individual-machine thinking was later developed by Stanton in Refs. \cite{5}-\cite{7}. In these studies, the individual-machine transient energy was developed to perform a machine-by-machine analysis \cite{5}.
Conjectures about the individual-machine potential energy surface were given in Ref. \cite{6}. The application of the individual-machine method in actual industrial environment was also demonstrated \cite{7}. The discussions, conjectures and also hypotheses in these papers greatly inspire the future individual machine studies.
\par Individual-machine method was initially proved to be effective. However, since the conjectures and hypotheses remain unsolved, this distinctive idea was at a standstill for decades. Recently, systematical progress was shown in the individual-machine studies.
The novel hybrid individual-machine-EAC method (IMEAC) was proposed \cite{8}-\cite{10}. The effectiveness of the method is explained through the individual-machine-transient-energy (IMTE) \cite{11}. Later, the reasonability of the IMTE is validated through the precise modeling of the individual-machine potential energy surface (IMPES) \cite{12}.
Further, the mapping between the generalized Newtonian system and multi-machine power system is established \cite{13}. Because strict mappings are found between the two-ball based Newtonian-system and two-machine system, the Newtonian energy conversion is successfully used as the stability evaluation of the multi-machine power system.
\par Because the individual-machine is physically real in a multi-machine power system, the ``machine stability” is naturally seen as the fundamental component for the multi-machine power system transient stability. Under this background, the analysis about the transient characteristics of the individual-machine becomes of value, because it can be used in the analysis of the mechanisms of the the superimposed-machine and equivalent-machine that are fully based on the ``energy superimposition” and ``motion equivalence” of the physically real individual-machines.

\subsection{SCOPE AND CONTRIBUTION OF THE PAPER}

In this paper, the mechanisms of the individual-machine method are revisited first. It is clarified that the individual-machine strictly follows the machine paradigms. These strict followings bring the advantages of the individual-machine method in TSA:
(i) the individual-machine trajectory stability is characterized precisely (stability characterization advantage), and (ii) the individual-machine trajectory variance is depicted clearly at the individual-machine MPP (trajectory-depiction advantage). The two advantages are fully reflected in the precise definitions of transient stability concepts. In particular, the swing of the critical machine is clearly depicted through its corresponding IMPP (trajectory-depiction advantage), the critical stability of the original system is strictly decided through the critical stability of the most-severely disturbed machine (the two advantages), and the he individual-machine potential energy surface is precisely modeled through the IMPE of the machine (the stability-characterization advantage).
Simulation results show that the strict followings of the paradigms fully ensure the effectiveness of the individual machine method in TSA.
\par The contributions of this paper are summarized as follows:
\\ (i) The individual-machine thinking comes from the monitoring of the trajectory variance of each physically real machine. This explains the modeling of the individual-machine.
\\ (ii) The individual-machine transient stability is established based on the machine paradigms. This provides a precise modeling and stability characterization for the individual machine.
\\ (iii) The transient stability concepts can be defined strictly through the individual machine. This provides the fundamental transient description of the original system trajectory.
\par The reminder of the paper is organized as follows. In Section \ref{section_II}, the mechanisms of the individual-machine are analyzed. In Section \ref{section_III}, the individual-machine based original system stability is given through the strict followings of the machine paradigms. In Section \ref{section_IV} the advantages of the individual-machine based transient stability concepts are analyzed. In Section \ref{section_V}, simulation cases show the effectiveness of the individual-machine in TSA. Conclusions are given in Section \ref{section_VI}.

\section{MECHANISMS OF THE INDIVIDUAL MACHINE}  \label{section_II}
\subsection{CRITICAL MACHINE MONITORING}  \label{section_IIA}

In the individual-machine based TSA, the system engineer monitors each critical machine in the original system trajectory \cite{14}.
\par The equation of motion of each physically real machine in the synchronous reference is denoted as
\begin{equation}
  \label{equ1}
  \left\{\begin{array}{l}
    \frac{d \delta_{i}}{d t}=\omega_{i} \\
    \\
    M_{i} \frac{d \omega_{i}}{d t}=P_{m i}-P_{e i}
    \end{array}\right.
\end{equation}
\par In Eq. (\ref{equ1}), all the parameters are given in Ref. \cite{8}.
\par After that, the equivalent Machine-SYS is used as the RM. The equation of motion of Machine-SYS is denoted as
\begin{equation}
  \label{equ2}
  \left\{\begin{array}{l}
    \frac{d \delta_{\mathrm{SYS}}}{d t}=\omega_{\mathrm{SYS}} \\
    \\
    M_{\mathrm{SYS}} \frac{d \omega_{\mathrm{SYS}}}{d t}=P_{\mathrm{SYS}}
    \end{array}\right.
\end{equation}
where
\begin{spacing}{1.5}
  \noindent$M_{\mathrm{SYS}}=\sum_{i=1}^{n} M_{i}$\\
  $\delta_{\mathrm{SYS}}=\frac{1}{M_{\mathrm{SYS}}} \sum_{i=1}^{n} M_{i} \delta_{i}$\\
  $\omega_{\mathrm{SYS}}=\frac{1}{M_{T}} \sum_{i=1}^{n} M_{i} \omega_{i}$\\
  $P_{\mathrm{SYS}}=\sum_{i=1}^{n} (P_{mi}-P_{ei})$
\end{spacing}
From Eq. (\ref{equ2}), the motion of Machine-SYS represents the equivalent motion of all machines in the system.
\par In the COI-SYS reference, the individual-machine trajectory (IMTR) is denoted as
\begin{equation}
  \label{equ3}
  \delta_{i\mbox{-}\mathrm{SYS}}=\delta_i-\delta_{\mathrm{SYS}}
\end{equation}
\par Based on Eq. (\ref{equ3}), the characteristics of the original system trajectory are given as below\\
\\ (i) The equivalent Machine-SYS is set as the RM.
\\ (ii) Using the Machine-SYS as the motion reference, the original system trajectory is formed by the IMTR of each physically real machine.
\\
\par According to (i) and (ii), the IMTR of each physically real machine (in the COI-SYS reference) is still preserved in the original system trajectory. Therefore, the original system trajectory is a multi-machine system trajectory.
\par
\begin{figure}[H]
  \centering
  \includegraphics[width=0.45\textwidth,center]{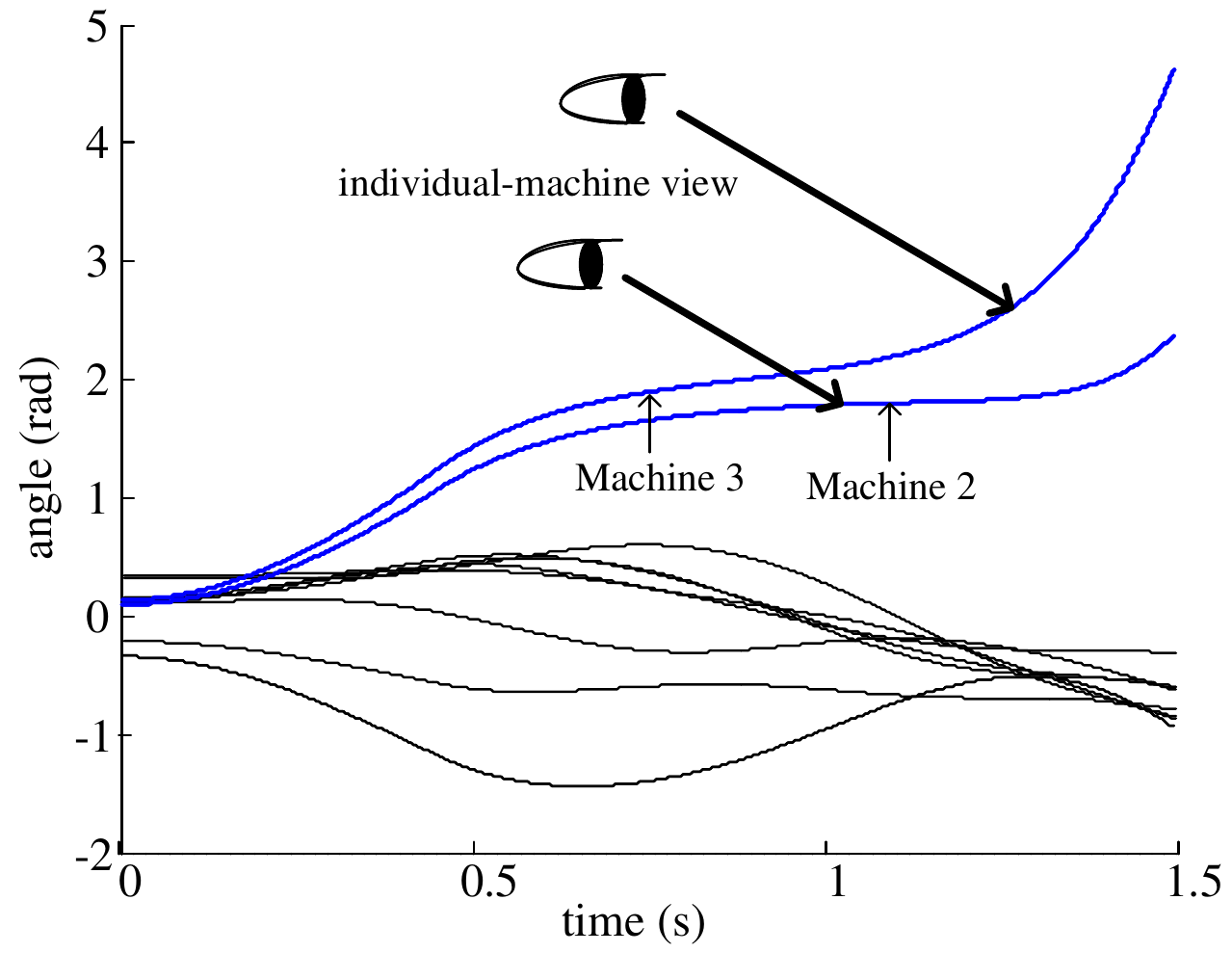}
  \caption{Original system trajectory in the COI-SYS reference [TS-1, bus-4, 0.447 s].} 
  \label{fig1}   
\end{figure}

 A tutorial example is given below to demonstrate the original system trajectory. The original system trajectory in the COI-SYS reference is shown in Fig. \ref{fig1}. In this case Machines 2 and 3 are critical machines because the two machines are severely disturbed.

\par In the transient stability analysis, a commonly observed phenomenon is that the IMTR of each ``critical machine” fluctuates severely, and thus it is most likely to separate from the system after fault clearing.
Comparatively, the IMTR of each non-critical machine fluctuates slightly and it hardly separate from the system.
Following trajectory stability theory, the system engineer may only monitor each critical machine by neglecting all the the non-critical machines. Furthermore, the prime objective of the system engineer is to monitor the stability of each critical machine in the system, because only the instability of each critical machine may cause the instability of the system.

\subsection{I-SYS SYSTEM MODELING} \label{section_IIB}
\begin{figure}[H]
  \centering
  \includegraphics[width=0.45\textwidth,center]{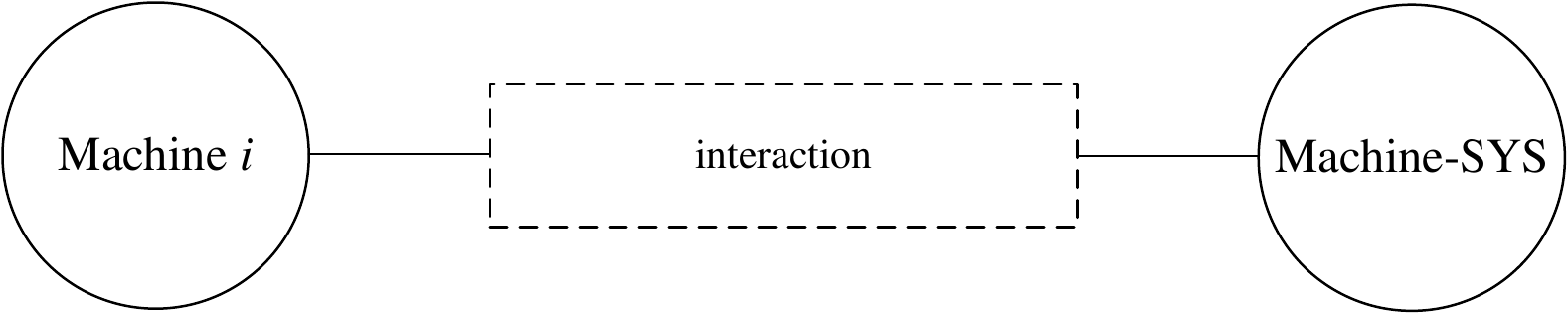}
  \caption{Formation of the I-SYS system.} 
  \label{fig2}   
\end{figure}

Based on the critical machine monitoring, the variance of the $\delta_{i\mbox{-}\mathrm{SYS}}$ of each critical machine is modeled through the I-SYS system \cite{8}. The formation of the I-SYS system is shown in Fig. \ref{fig2}.
Based on Eqs. (\ref{equ1}) and (\ref{equ2}), the relative motion between Machine \textit{i} and Machine-SYS is depicted as
\begin{equation}
  \label{equ4}
  \left\{\begin{array}{l}
    \frac{d \delta_{i\mbox{-}\mathrm{SYS}}}{d t}=\omega_{i\mbox{-}\mathrm{SYS}} \\
    \\
    M_{i} \frac{d \omega_{i\mbox{-}\mathrm{SYS}}}{d t}=f_{i\mbox{-}\mathrm{SYS}}
    \end{array}\right. 
   \end{equation}
where
\begin{spacing}{1.5}
  \noindent $f_{i\mbox{-}\mathrm{SYS}}=P_{m i}-P_{e i}-\frac{M_{i}}{M_{\text {SYS}}} P_{\mathrm{SYS}}$ \\
  $\omega_{i\mbox{-}\mathrm{SYS}}=\omega_{i}-\omega_{\mathrm{SYS}}$
\end{spacing}
The individual-machine DLP (IDLP) is denoted as
\begin{equation}
  \label{equ5}
  f_{i\mbox{-}\mathrm{SYS}}=0
\end{equation}
\par In Eq. (\ref{equ5}), the IDLP of Machine \textit{i} depicts the point where the machine becomes unstable.

\subsection{INDIVIDUAL-MACHINE TRANSIENT ENERGY CONVERSION} \label{section_IIC}

The IMTE is defined in a typical Newtonian energy manner. The IMTE of the machine is defined as
\begin{equation}
  \label{equ6}
  V_{i\mbox{-}\mathrm{SYS}}=V_{K E i\mbox{-}\mathrm{SYS}}+V_{P E i\mbox{-}\mathrm{SYS}}
\end{equation}
where
\begin{spacing}{1.5}
 \noindent$V_{K E i\mbox{-}\mathrm{SYS}}=\frac{1}{2} M_{i} \omega_{i\mbox{-}\mathrm{SYS}}^{2}$\\ 
 $V_{P E i\mbox{-}\mathrm{SYS}}=\int_{\delta_{i\mbox{-}\mathrm{SYS}}^{s}}^{\delta_{i\mbox{-}\mathrm{SYS}}}\left[-f_{i\mbox{-}\mathrm{SYS}}^{(P F)}\right] d \delta_{i\mbox{-}\mathrm{SYS}}$
\end{spacing}
In Eq. (\ref{equ6}), the conversion between IMKE and IMPE is used to measure the stability of the I-SYS system.
\par  The residual KE of Machine \textit{i} at its corresponding MPP is denoted as
\begin{equation}
  \label{equ7}
  \begin{split}
    V_{K E i\mbox{-}\mathrm{SYS}}^{R E}&=V_{K E i\mbox{-}\mathrm{SYS}}^{c}-\Delta V_{P E i\mbox{-}\mathrm{SYS}} \\
    &=A_{A C C i\mbox{-}\mathrm{SYS}}-A_{D E C i\mbox{-}\mathrm{SYS}} 
  \end{split} 
\end{equation}
where
\begin{spacing}{2}
  \noindent$V_{K E i\mbox{-}\mathrm{SYS}}^{c}=\frac{1}{2} M_{i} \omega_{i\mbox{-}\mathrm{SYS}}^{c 2}=A_{A C C i\mbox{-}\mathrm{SYS}}$\\
  \noindent$\Delta V_{P E i\mbox{-}\mathrm{SYS}}=\int_{\delta_{i\mbox{-}\mathrm{SYS}}^{s}}^{\delta_{i\mbox{-}\mathrm{SYS}}^{MPP}}\left[-f_{i\mbox{-}\mathrm{SYS}}^{(P F)}\right] d \delta_{i\mbox{-}\mathrm{SYS}}-\\
  \int_{\delta_{i\mbox{-}\mathrm{SYS}}^{s}}^{\delta_{i\mbox{-}\mathrm{SYS}}^{c}}\left[-f_{i\mbox{-}\mathrm{SYS}}^{(P F)}\right] d \delta_{i\mbox{-}\mathrm{SYS}}
  =A_{DECi\mbox{-}\mathrm{SYS}}$
\end{spacing}
In Eq. (\ref{equ7}), note that the individual-machine transient energy conversion is identical to the IMEAC \cite{11}.
\par The stability characterizations of the individual machine are summarized as below.\\
\\ (i) From the perspective of transient energy conversion, the individual machine is evaluated to go unstable if the residual IMKE occurs at its IMPP.
\\ (ii) From the perspective of EAC, the individual machine is evaluated to go unstable if the acceleration area is larger than the deceleration area.
\\
\par
\begin{figure}[H]
  \centering
    \subfigure[]{  
      \includegraphics[width=0.45\textwidth,center]{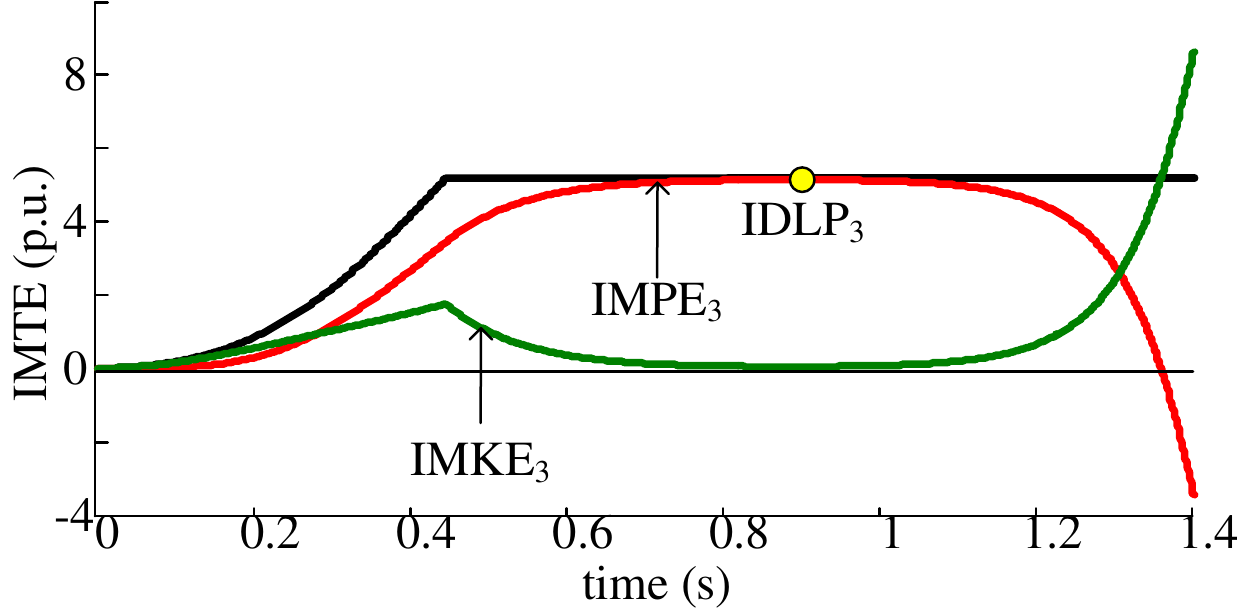}}    
      \centering
      \subfigure[]{
        \includegraphics[width=0.45\textwidth,center]{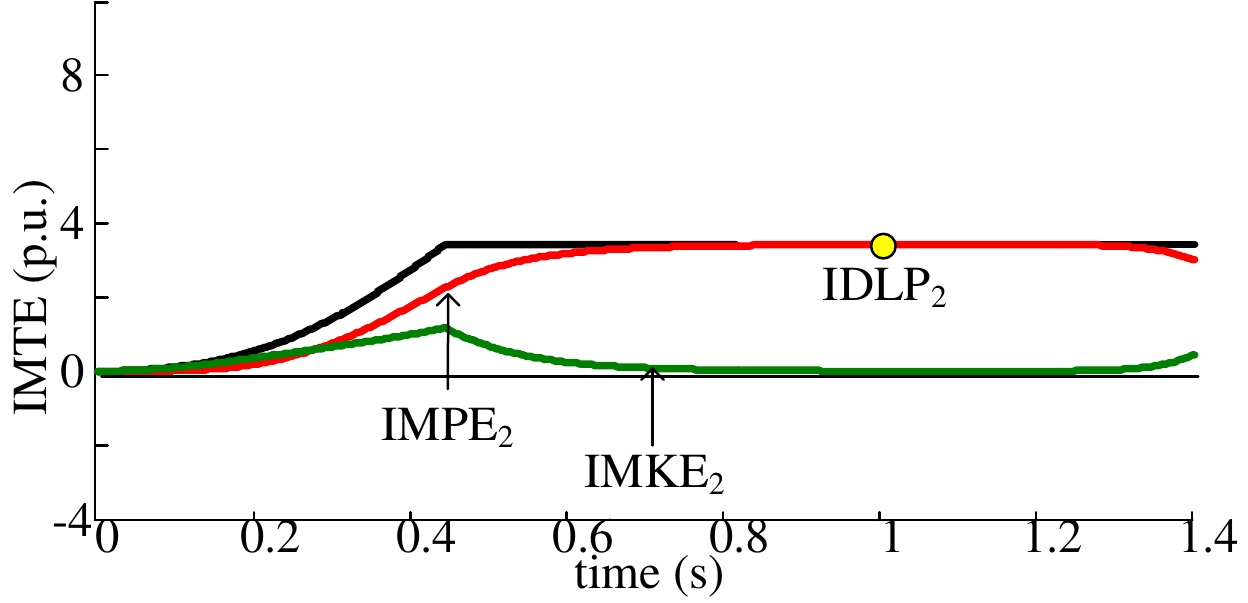}}     
   \caption{Individual machine transient energy conversion [TS-1, bus-4, 0.447 s]. (a) Machine 3. (b) Machine 2.}
  \label{fig3}
\end{figure}

 For the case as in Fig. \ref{fig1}, the individual-machine transient energy conversion inside each critical machine is shown in Fig. \ref{fig3}.
The IMEAC of each machine is shown in Fig. \ref{fig4}. Note that the individual-machine transient energy conversion is completely identical to the IMEAC. 
The two are just the different expressions of the energy conversion in the \textit{t-V} space and $\delta\mbox{-}f$ space, respectively.

\begin{figure} [H]
  \centering 
  \subfigure[]{%
  \label{fig4a}
    \includegraphics[width=0.37\textwidth]{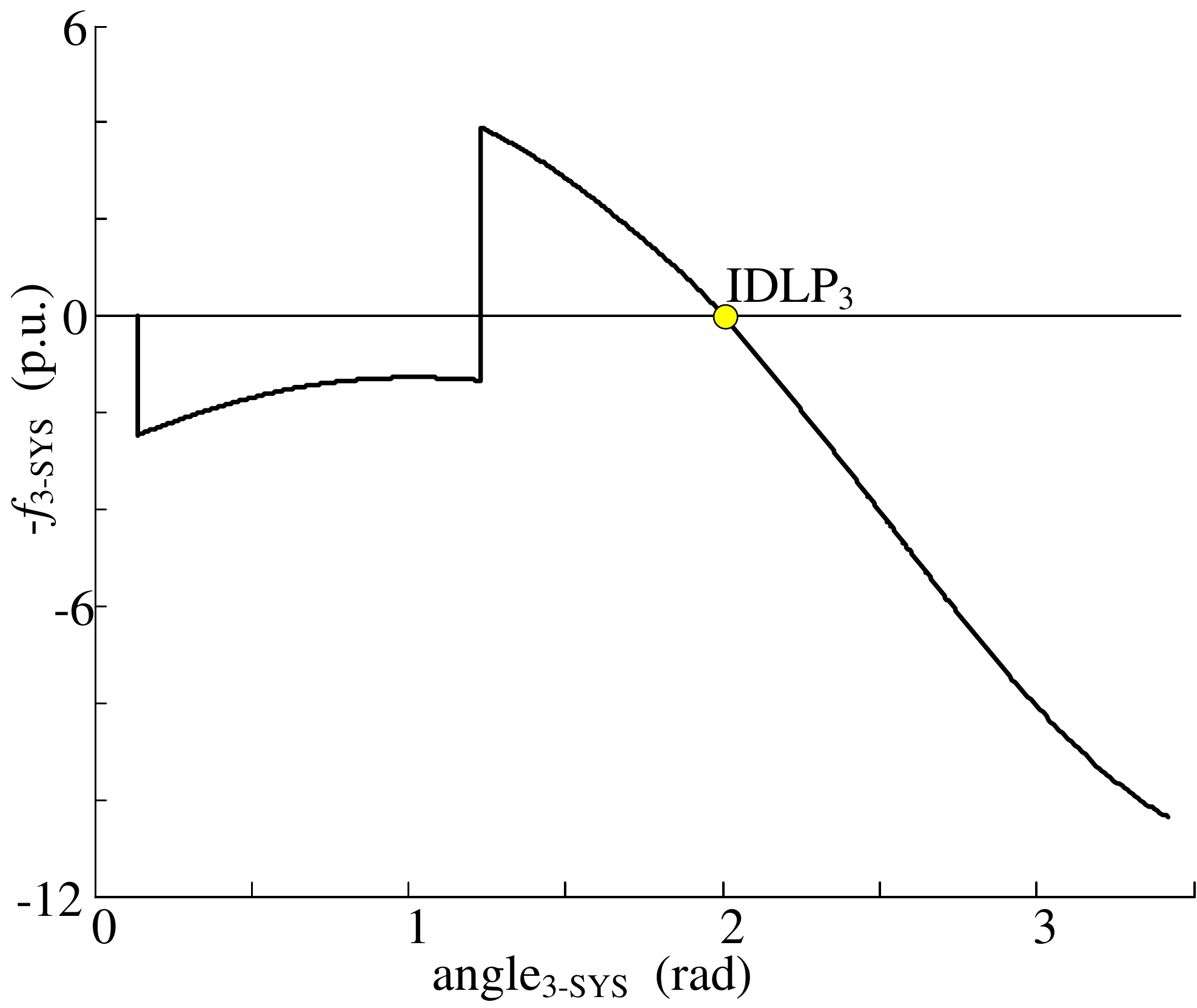}}%

\end{figure} 
\addtocounter{figure}{-1}       
\begin{figure} [H]
  \addtocounter{figure}{1}      
  \centering 
  \subfigure[]{%
    \label{fig4b}
    \includegraphics[width=0.37\textwidth]{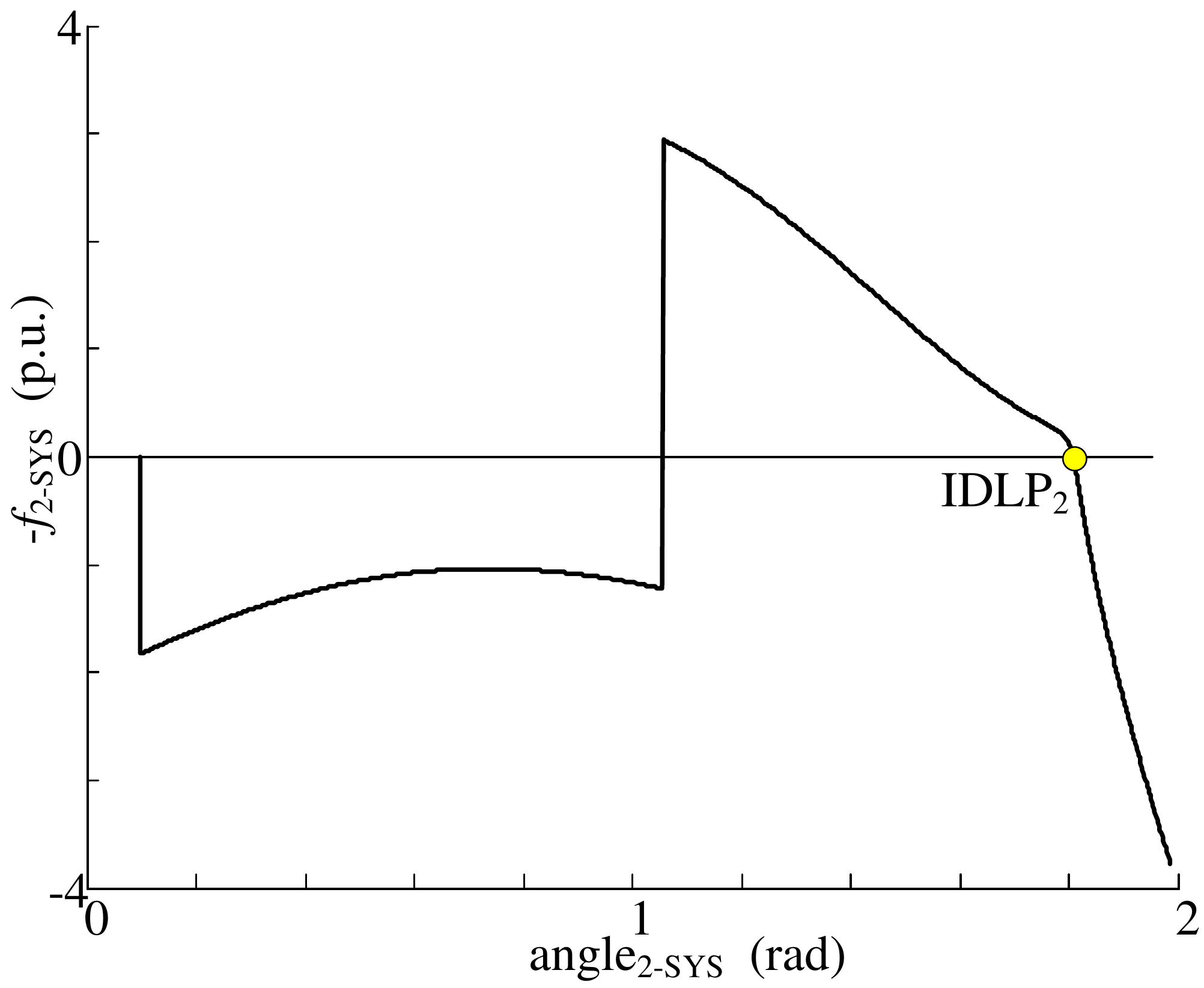}}%
  \caption{IMEAC [TS-1, bus-4, 0.447 s]. (a) Machine 3. (b) Machine 2.}%
  \label{fig4}
\end{figure}

\section{INDIVIDUAL MACHINE BASED ORIGINAL SYSTEM STABILITY} \label{section_III}
\subsection{FOLLOWINGS OF THE MACHINE PARADIGMS} \label{section_IIIA}
Based on the analysis in Section \ref{section_II}, the individual-machine strictly follows all the paradigms.
\\
\par\noindent\textit{Following of the trajectory paradigm}: From Section \ref{section_IIA}, the system engineer monitors the IMTR of each machine in the COI-SYS reference, i.e., the separation of the machine with respect to the Machine-SYS. Therefore, this IMTR monitoring strictly follows the trajectory paradigm. 
\\ \textit{Following of the modeling paradigm}: From Section \ref{section_IIB}, the IMTR of each machine is modeled through its corresponding I-SYS system that is formed by the physically real machine and Machine-SYS. Therefore, this I-SYS system modeling strictly follows the modeling paradigm.
\\ \textit{Following of the energy paradigm}: From Section \ref{section_IIC}, the IMTE is defined in an Newtonian energy manner. Therefore, this Newtonian definition of the IMTE strictly follows the energy paradigm. The IMEAC also holds in the individual-machine.
\\
\par The strict followings of the machine paradigms will fully ensure the advantages of the individual machine in TSA. This will be analyzed in the following section.

\subsection{MACHINE STABILITY AND SYSTEM STABILITY} \label{section_IIIB}
\noindent \textit{Critical machine stability}: The stability margin of each critical machine is denoted as
\begin{equation}
  \label{equ8}
  \eta_{i}=\frac{\Delta V_{P E i\mbox{-}\mathrm{SYS}}-V_{K E i\mbox{-}\mathrm{SYS}}^{c}}{V_{\mathrm{KE} i\mbox{-}\mathrm{SYS}}^{c}}=\frac{A_{\text {dec} i\mbox{-}\mathrm{SYS}}-A_{\mathrm{acc} i\mbox{-}\mathrm{SYS}}}{A_{\text {acc} i\mbox{-}\mathrm{SYS}}}
\end{equation}
where
\begin{spacing}{2}
  \noindent$V_{K E i\mbox{-}\mathrm{SYS}}^{c}=A_{\mathrm {acc} i \mbox{-} \mathrm{SYS} }=\frac{1}{2} M_{i} \omega_{i\mbox{-}\mathrm{SYS}}^{c 2}$\\
  $\Delta V_{P E i\mbox{-}\mathrm{SYS}}=A_{\mathrm {dec} i \mbox{-} \mathrm{SYS}}=
  \int_{\delta_{i\mbox{-} \mathrm{SYS}}^{c}}^{\delta_{i\mbox{-}\mathrm{SYS}}^{IMPP}}\left[-f_{i\mbox{-}\mathrm{SYS}}^{(P F)}\right] \,d \delta_{i\mbox{-}\mathrm{SYS}}$
\end{spacing}
From Eq. (\ref{equ8}), the stability state of the machine can be characterized through the sign of $\eta_{i} $: $\eta_{i}>0$ means that the machine is stable;
$\eta_{i}=0$ means that the machine is critical stable; and $\eta_{i}<0$ means that the machine becomes unstable; The stability margin, i.e., the ``severity'' of the machine is measured through the absolute value of $\eta_i$.
\\
\textit{Original system stability}:  For the multi-machine system with \textit{n} machines, because the IMTR of each machine is preserved in the original system trajectory, the stability of the entire system is decided by the stability of each machine. Under this background, the stability margin of the original system is defined as
\begin{equation}
  \label{equ9}
  \eta_{\mathrm{sys}}=\left[\eta_i\right] 
\end{equation}
\par From Eq. (\ref{equ9}), $\eta_{\mathrm{sys}}$ comprise of the stability margin of each machine because the original system trajectory is preserved in TSA.
\\
\textit{Machine-by-machine stability evaluation}: In the original system, the stability of each machine is monitored and characterized independently in parallel in TSA. Since the NEC inside each machine is also unique and different, the IDSP or IDLP of the machine will occur one after another, and thus the machine will be characterized as maintaining stable or becoming unstable one after another. This fully indicates that the original multi-machine system should be evaluated in a ``machine-by-machine” manner.
\par According to the unity principle, the ``machine-by-machine” stability characterization of the original system is given as below
\\
\\ (i) The instability of the system is identified once the ``first” IDLP (the IDLP of the leading unstable machine) occurs along time horizon.
\\ (ii) The severity of the system is obtained only when the ``last” IDLP occurs along time horizon.
\\
\par (i) and (ii) indicate that the stability and the severity of the system will be obtained at different time points in the individual-machine based TSA. Detailed analysis is given in the case study.

\subsection{ADVANTAGES OF THE INDIVIDUAL ACHINE IN TSA} \label{section_IIIC}
Revisiting the analysis in Section \ref{section_IIIA}, the individual-machine analyst focuses on the transient characteristic of each critical machine in the original system trajectory. Based on this individual-machine monitoring, the I-SYS system of each machine is established with strict NEC characteristic. Further, extending to the system level, because the original system is formed by multiple machines, each critical machine is monitored in parallel, and the stability of each critical machine is characterized independently \cite{13}.
\par
\begin{figure}[H]
  \centering
  \includegraphics[width=0.45\textwidth,center]{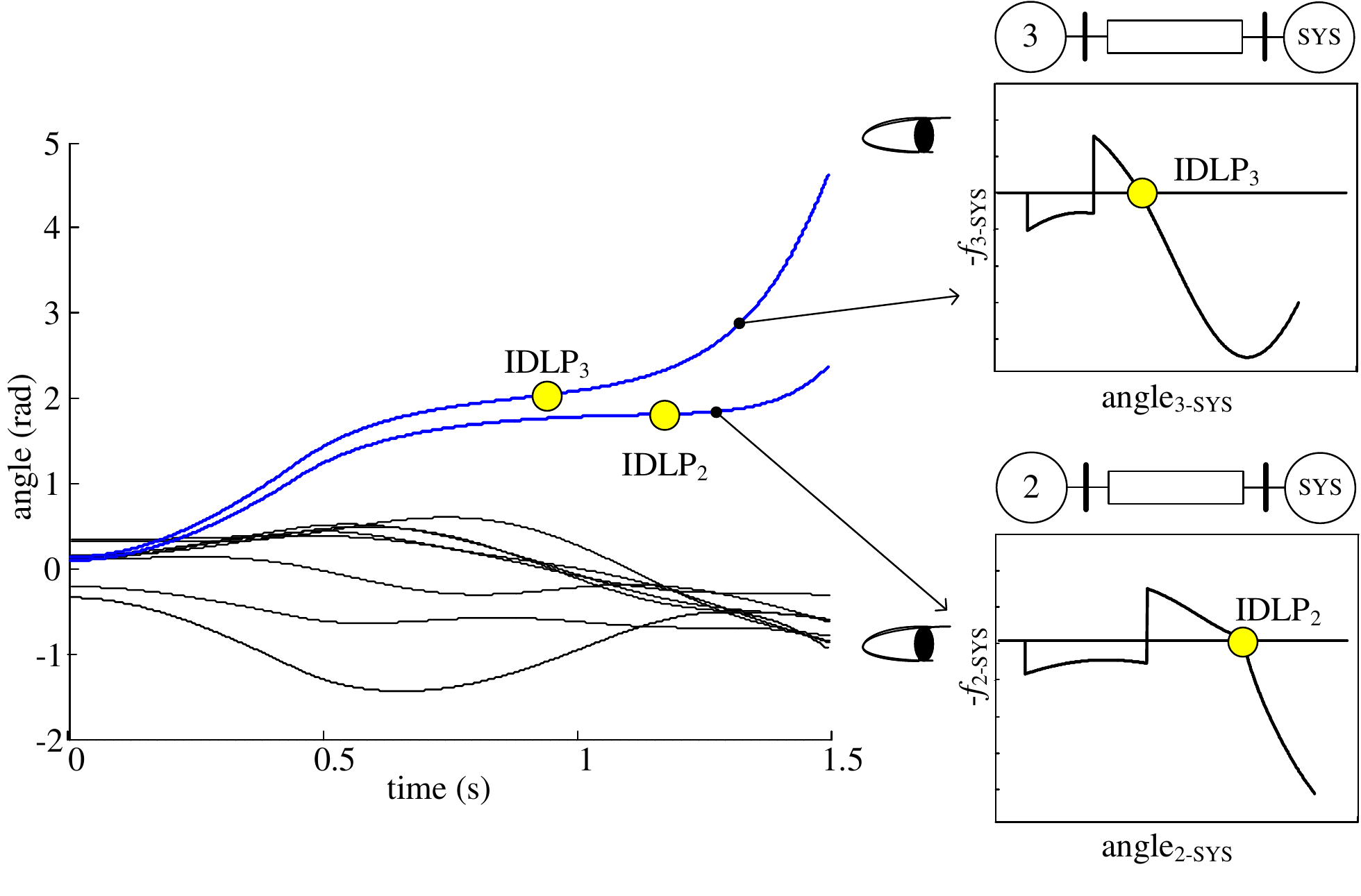}
  \caption{The use of the individual-machine in the original system [TS-1, bus-4, 0.447 s].} 
  \label{fig5}   
\end{figure}

 The use of the individual machine in the original system is shown in Fig. \ref{fig5}.
The strict followings of the machine paradigms essentially ensure the advantages of the individual-machine and also its application in TSA. The advantages are given as below:
\\
\\ Stability-characterization advantage: The stability of IMTR of each critical machine is characterized precisely at IMPP.
\\Trajectory-depiction advantage: The variance of IMTR of each critical machine is depicted clearly through the individual-machine IMPP.
\\
\par The two advantages are fully based on the strict correlation between the trajectory variance and the transient energy conversion through the I-SYS system modeling \cite{14}.
The two advantages will be fully reflected in the definitions of the individual-machine based transient stability concepts. Detailed analysis is given in the following section.

\section{PRECISE DEFINITIONS OF THE INDIVIDUAL-MACHINE BASED TRANSIENT STABILITY CONCEPTS} \label{section_IV}
\subsection{CRITICAL MACHINE SWING} \label{section_IVA}
\noindent \textit{Statement}: The trajectory-depiction advantage is fully reflected in the definition of the critical-machine swing.
\\ \textit{Individual-machine perspective}: The system engineer monitors the IMTR of each critical machine in the original system. The definitions of the stable and unstable critical-machine swing are given as below
\\ Swing of a stable critical machine: It is defined as the IDSP of the machine. This is because IDSP reflects the ``inflection” of the IMTR of the machine $\left(\mathrm{d} \delta_{i\mbox{-}\mathrm{SYS}} / \mathrm{d} t=\omega_{\mathrm{IDSP}}=0\right) $.
\\ Swing of an unstable critical machine: It is defined as the IMDLP of the machine. This is because IDLP reflects the ``separation” of the IMTR of the machine $\left(\mathrm{d}^{2} \delta_{i\mbox{-}{\mathrm{SYS}}}/ \mathrm{d} t^2= f_{\mathrm{IDLP}}=0\right)$.
\par
\begin{figure}[H]
  \centering
  \includegraphics[width=0.4\textwidth,center]{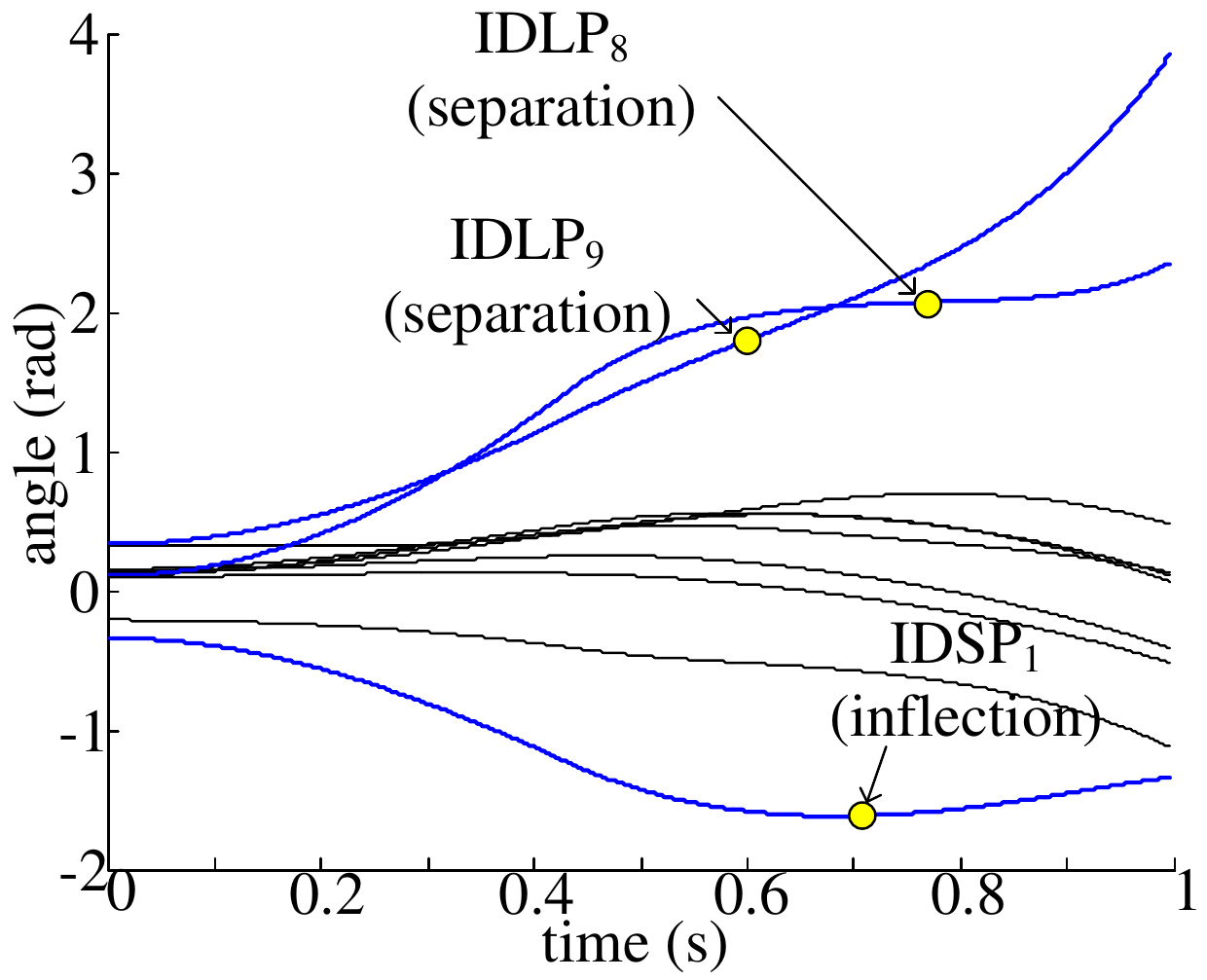}
  \caption{System trajectory [TS-1, bus-2, 0.430 s].} 
  \label{fig6}   
\end{figure}

\begin{figure}[H]
  \centering
  \includegraphics[width=0.4\textwidth,center]{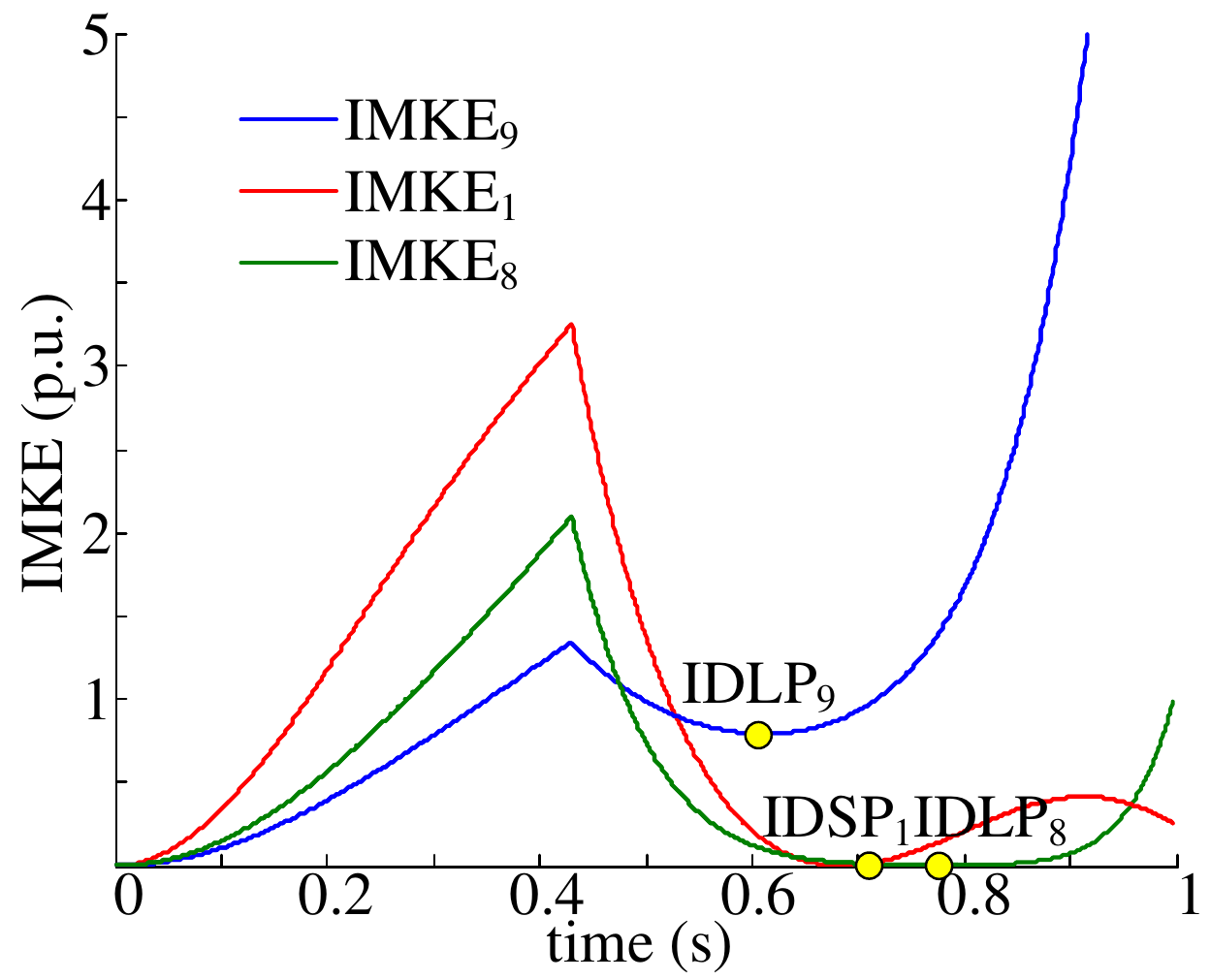}
  \caption{IMNEC inside each critical machine[TS-1, bus-2, 0.430 s].} 
  \label{fig7}   
\end{figure}

 Based on the definitions above, the concept of the critical machine swing focuses on the depiction of the trajectory variance of each critical machine in the original system.
\\ \textit{Example}: The concept of the critical-machine swing is demonstrated as below. In this case Machines 8, 9 and 1 are critical machines. Both Machines 8 and 9 go unstable. The system trajectory is shown in Fig. \ref{fig6}.
The individual-machine transient energy conversion of each critical machine is shown in Fig. \ref{fig7}.
\par From Figs. \ref{fig6} and \ref{fig7}, the variance of the IMTR of each critical machine is depicted as below
\\
\\ \textit{$IDLP_9$ occurs (0.614 s)}: $\text{IMTR}_9$ starts separating from the system at the moment. Machine 9 goes first-swing unstable. 
\\ \textit{$IDSP_1$ occurs (0.686 s)}: $\text{IMTR}_1$ inflects back at the moment. Machine 1 maintains first-swing stable.
\\ \textit{$IDLP_8$ occurs (0.776 s)}: $\text{IMTR}_8$ starts separating from the system at the moment. Machine 8 goes first-swing unstable.
\\
\par From analysis above, the variance of the IMTR of each critical machine is depicted clearly at IMPP.

\subsection{CRITICAL STABILITY OF THE ORIGINAL SYSTEM} \label{section_IVB}
\begin{figure}[H]
  \centering
    \subfigure[]{  
      \includegraphics[width=0.42\textwidth,center]{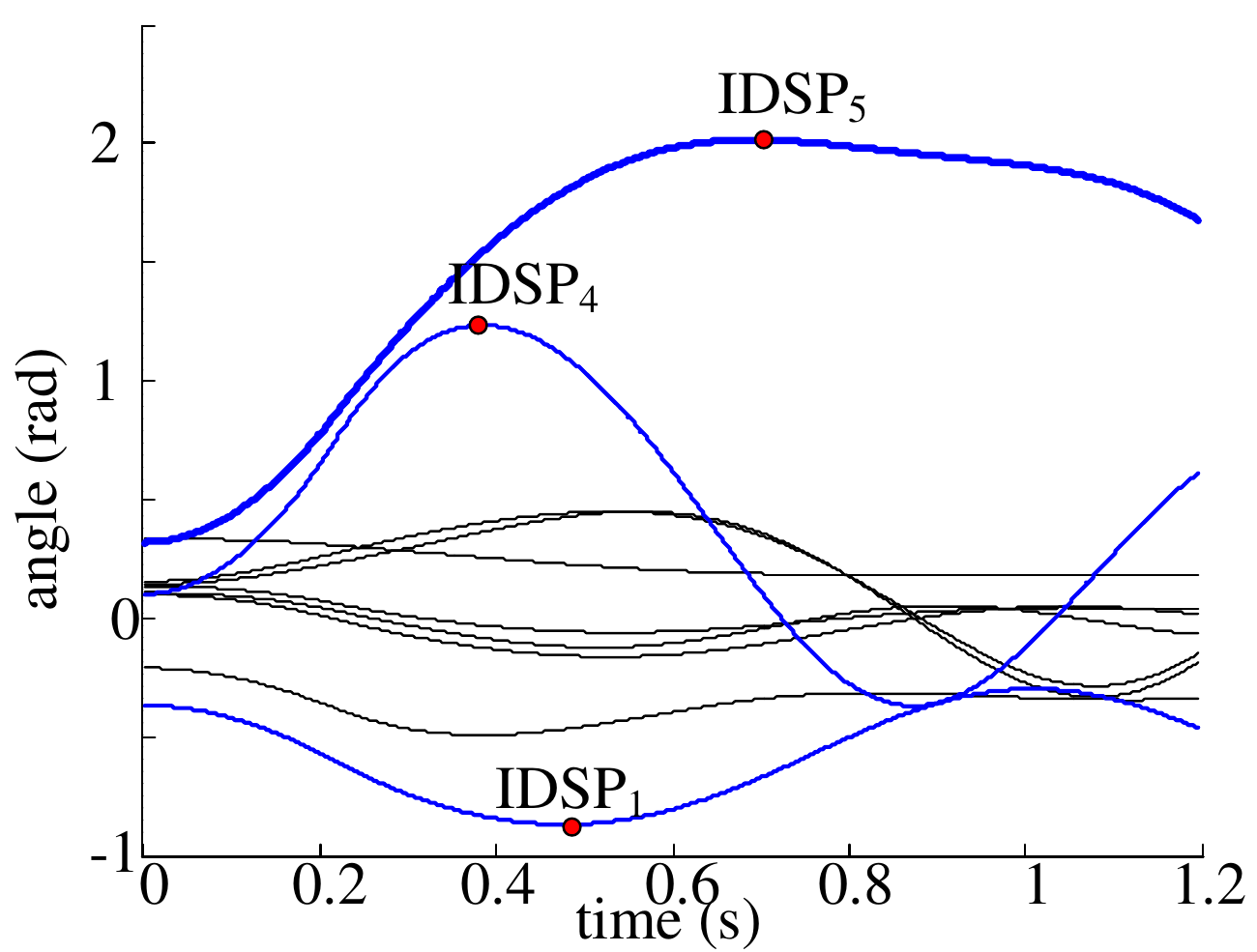}}    
      \centering
      \subfigure[]{
        \includegraphics[width=0.42\textwidth,center]{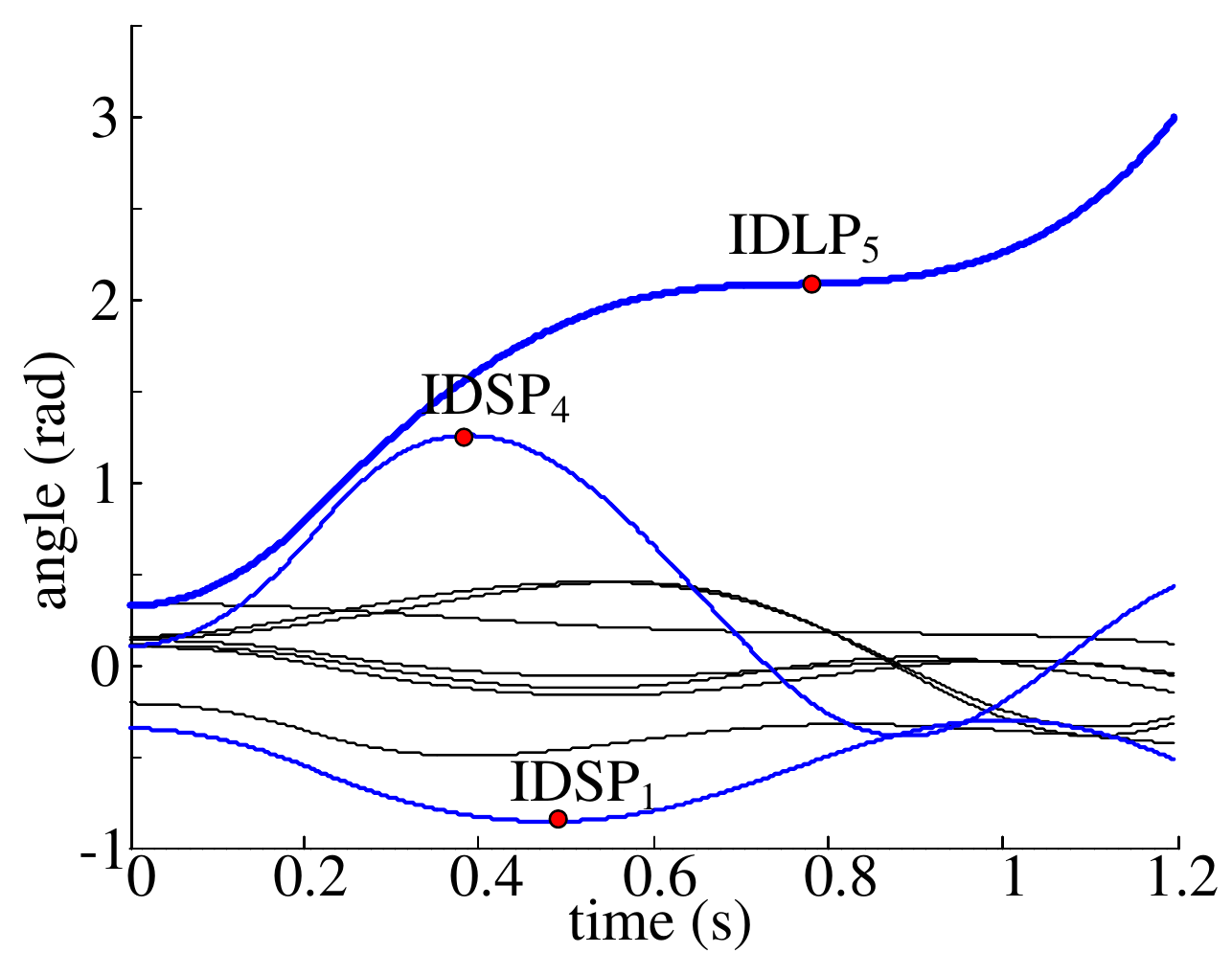}}     
   \caption{System trajectory. (a) Critical stable case [TS-1, bus-19, 0.215s]. (b) Critical unstable case [TS-1, bus-19, 0.216s].}
  \label{fig8}
\end{figure}

\noindent\textit{Statement}: Both the stability-characterization advantage and the trajectory-depiction advantage are reflected in the definition of the critical stability of the original system.
\\ \textit{Individual-machine perspective}: The critical stability state of the original system is completely decided through the critical stability of the most-severely disturbed machine (MDM). Therefore, the IDSP of the critical stable MDM becomes crucial in the critical stability analysis of the original system. 
\\ \textit{Example}: The simulation case of the critical stable original system trajectory is given as below. The system trajectory is defined in the COI-SYS reference. The CCT is 0.215 s. 
Machine 5 is also the MDM in this case \cite{9}, \cite{11}. The system trajectories being critical stable and critical unstable are shown in Figs. \ref{fig8} (a) and (b), respectively. The IMEAC inside the MDM is shown in Fig. \ref{fig9}.

\begin{figure} [H]
  \centering 
  \subfigure[]{%
  \label{fig9a}
    \includegraphics[width=0.32\textwidth]{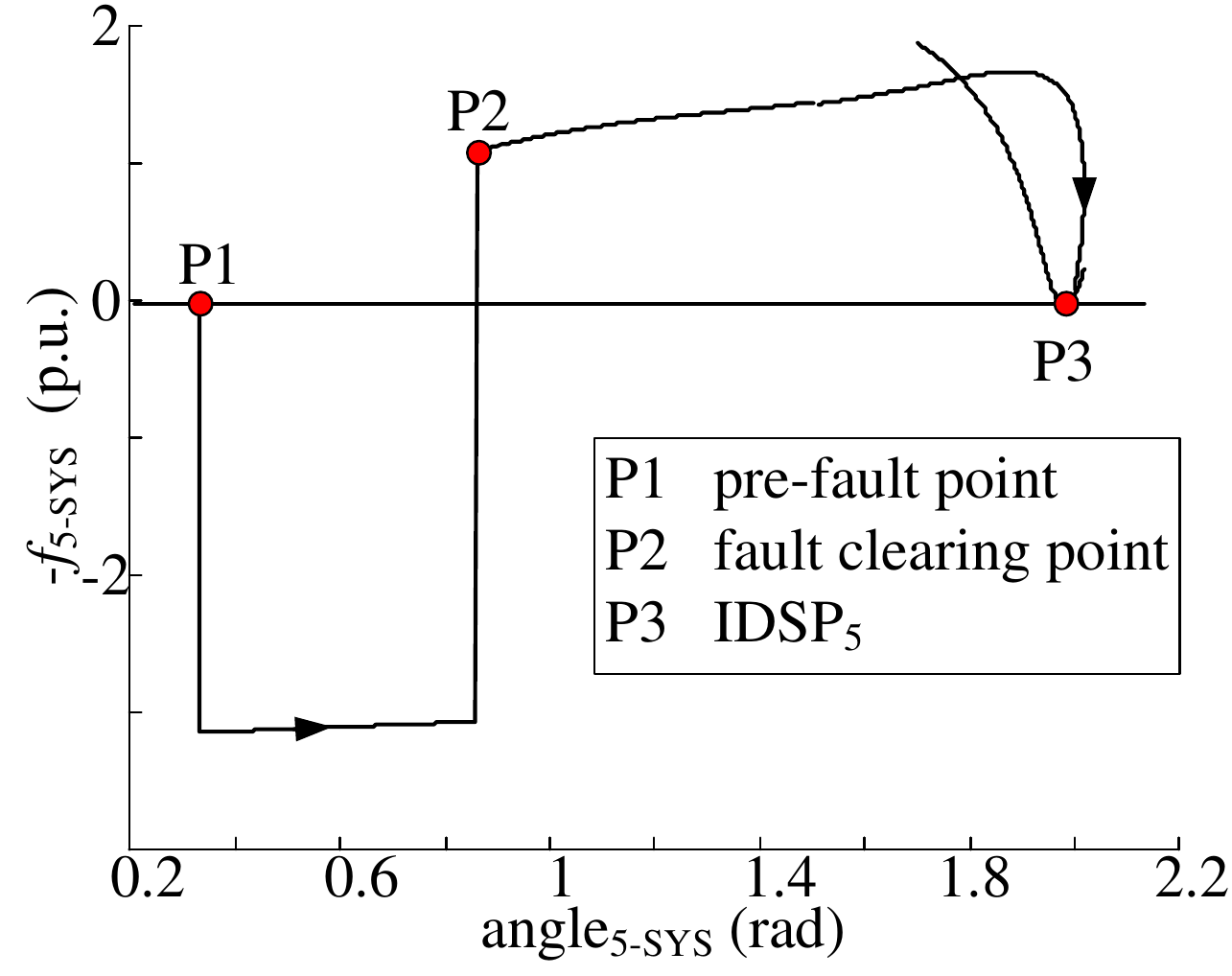}}%
\end{figure} 
\addtocounter{figure}{-1}       
\begin{figure} [H]
  \addtocounter{figure}{1}      
  \centering 
  \subfigure[]{%
    \label{fig9b}
    \includegraphics[width=0.32\textwidth]{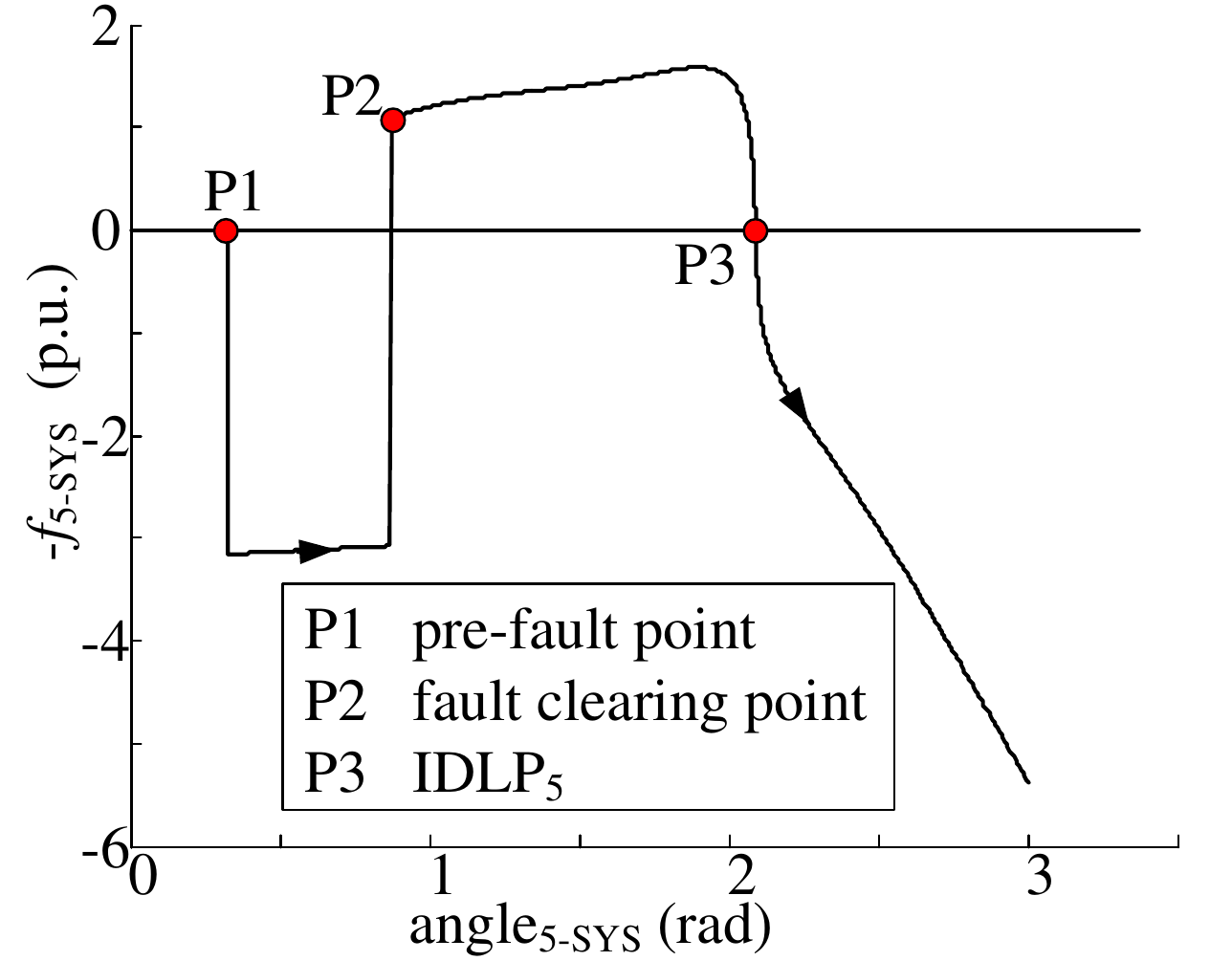}}%
  \caption{IMEAC of the MDM. (a) Critical stable case. (b) Critical unstable case.}%
  \label{fig9}
\end{figure}
\par The analysis of the critical stability of the original system is given as below
\\
\\ MDM monitoring: Through the comparison between the critical stable case and the critical unstable case as given in Figs. \ref{fig8} (a) and (b), the critical stability of the entire original system is completely decided by the critical stability of the MDM.
\\ I-SYS system modeling: According to the modeling paradigm, the trajectory stability of the MDM is modeled through the corresponding $\text{I}\mbox{-}\text{SYS}_{\text{MDM}}$ system.
\\ IMEAC: According to the energy paradigm, MDM maintains critical stable when $t_{\text{c}}$ is 0.215 s, as in Fig. \ref{fig9a}. MDM becomes critical unstable when $t_{\text{c}}$ is 0.216 s, as in Fig. \ref{fig9b}.
According to the unity principle, the critical stability of the system is completely decided by the MDM. 
\\
\par
\begin{figure}[h]
  \centering
  \includegraphics[width=0.36\textwidth,center]{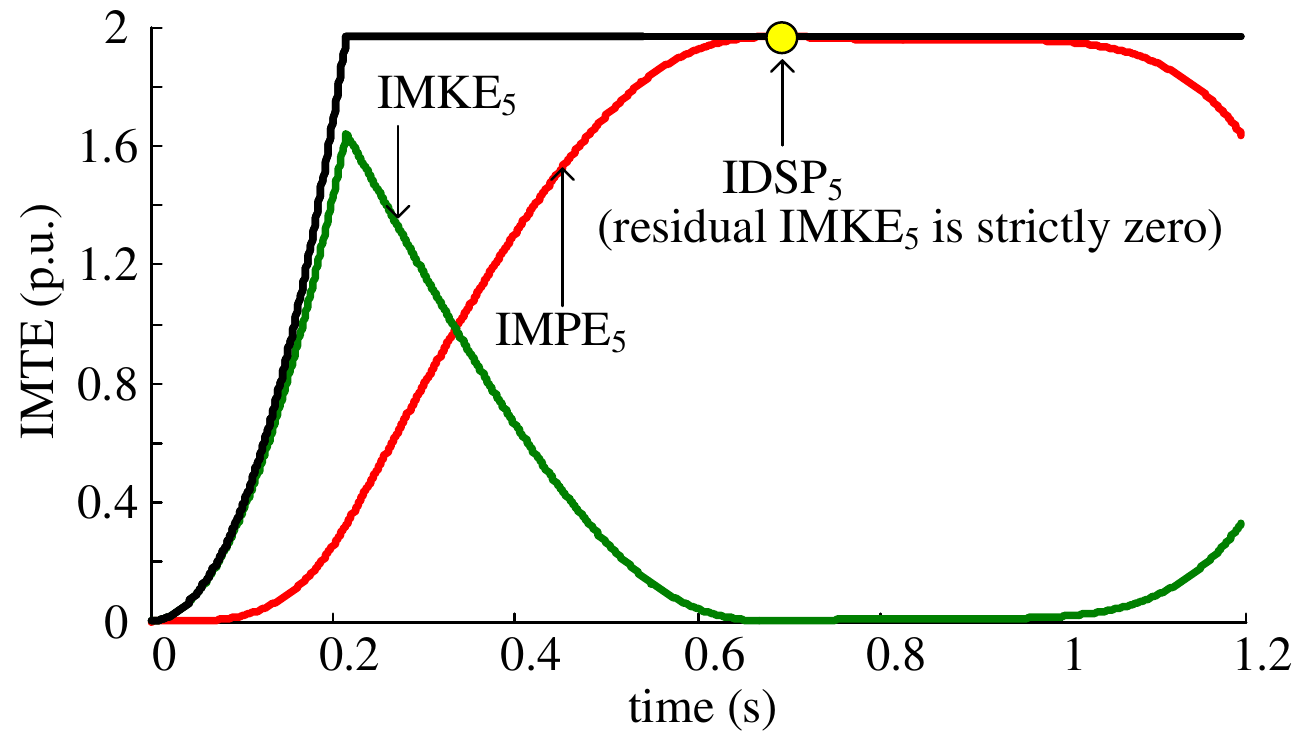}
  \caption{Transient energy conversion inside critical-stable Machine 5 [TS-1, bus-19, 0.215 s].} 
  \label{fig10}  
\end{figure}

 We go a further step. Based on the modeling of the $\text{I}\mbox{-}\text{SYS}_{\text{MDM}}$, the critical transient energy of the entire system is also defined as the critical transient energy of MDM. The individual-machine transient energy conversion inside Machine 5 is shown in Fig. \ref{fig10} \cite{11}.

\begin{table}[H]\scriptsize
  \centering
  \caption{$\text{IMPE}_\text{5}$ and $\text{IMKE}_\text{5}$ at $\text{IMPP}_\text{5}$} 
  \begin{tabular}{@{}cccc@{}}
  \toprule
  system state        & \begin{tabular}[c]{@{}c@{}}$\text{IMTE}_\text{5}$ at fault\\ clearing point\\ (p.u.)\end{tabular} & \begin{tabular}[c]{@{}c@{}}$\text{IMPE}_\text{5}$ at $\text{IMPP}_\text{5}$\\ (p.u.)\end{tabular} & \begin{tabular}[c]{@{}c@{}}Residual $\text{IMKE}_\text{5}$\\ (p.u.)\end{tabular} \\ \midrule
  Critically stable   & 1.9672                                                                           & 1.9672                                                          & {\ul 0.0000}                                                    \\
  Critically unstable & 2.0079                                                                           & 2.0073                                                          & 0.0006                                                          \\ \bottomrule
  \end{tabular}

  \label{table1}
\end{table}
From analysis above, the individual machine shows two advantages in the definitions of the critical stability of the original system.
\\
(i) The stability state of the MDM from the critical stability to the critical instability is characterized precisely through NEC, as in Figs. \ref{fig9}, \ref{fig10} and Table \ref{table1}.
\\
(ii) The trajectory variance of the MDM from the critical stability to the critical instability is depicted clearly through the change from $\text{IDSP}_{\text{MDM}}$ to the $\text{IDLP}_{\text{MDM}}$, as in Fig. \ref{fig8}.
\par (i) and (ii) are fully based on the strict followings of the machine paradigms in the individual machine.
\par
\begin{figure}[h]
  \centering
  \includegraphics[width=0.42\textwidth,center]{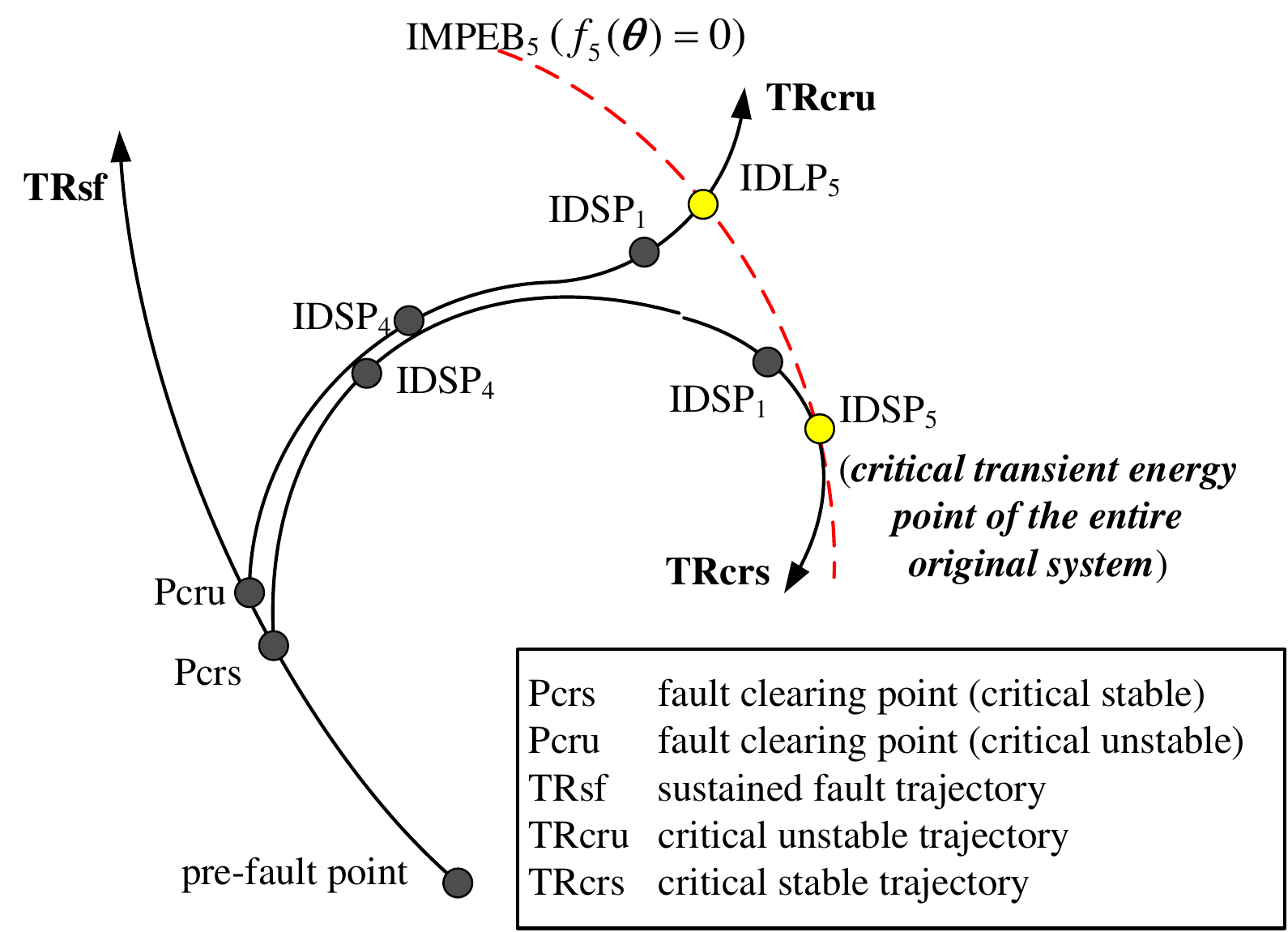}
  \caption{Critical energy point of the system from individual-machine angle.} 
  \label{fig11}  
\end{figure}

 Demonstration of the critical stability of the system from individual-machine perspective is shown in Fig. \ref{fig11}.

From Fig. \ref{fig11}, the critical energy point of the system i.e., $\text{IDSP}_{\text{5}}$ lies in the $\text{IMPEB}_{\text{5}}$. The stability boundary of the entire system is also defined as the $\text{IMPEB}_{\text{5}}$.
The critical stability of the system can be characterized precisely through the system trajectory going through $\text{IMPEB}_{\text{5}}$. Note that each machine corresponds to its unique IMPEB in a multi-machine system, yet only $\text{IMPEB}_{\text{5}}$ is shown in the figure for clearance.

\subsection{INDIVIDUAL-MACHINE POTENTIAL ENERGY SURFACE} \label{section_IVC}
\begin{figure}[H]
  \centering
  \includegraphics[width=0.42\textwidth,center]{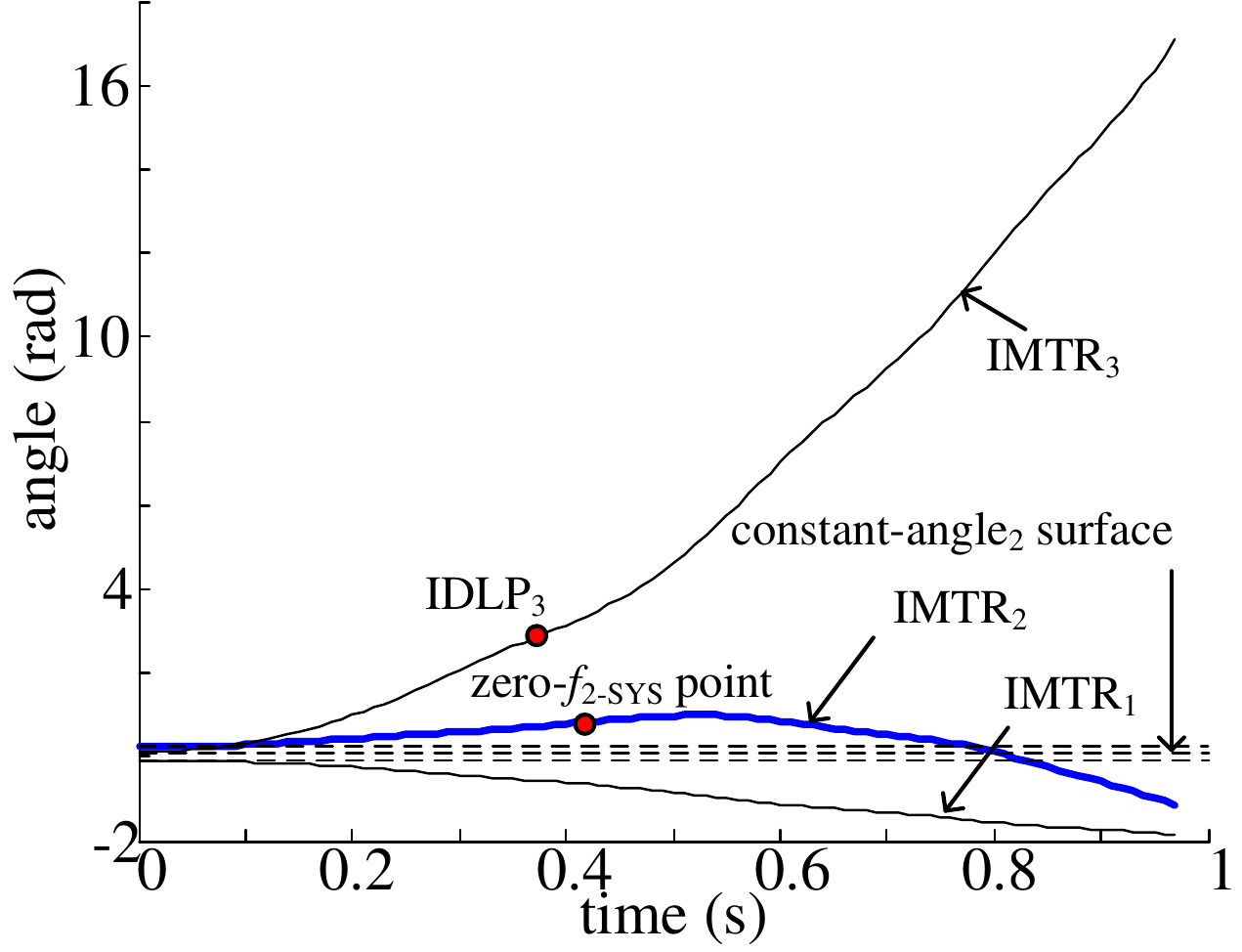}
  \caption{IMTR of Machine 2 [TS-4, bus-3, 0.30 s].} 
  \label{fig12}  
\end{figure}

\noindent \textit{Statement}: The stability characterization advantage is fully reflected in the precise modeling of the IMPES.
\\ \textit{Individual-machine perspective}: the IMPES is modeled completely based on the IMPE of each machine \cite{10}. In particular, assume numerous system trajectories in the angle space form a ``system-trajectory set”.
Each system trajectory in this set corresponds to the IMPE of Machine \textit{i} along it. Next, all these IMPEs of Machine \textit{i} under different system trajectories may form the IMPES of the machine.
\\ \textit{Example}: The IMTR of Machine 2 is shown in Fig. \ref{fig12}. Demonstrations about the motion of the energy ball on its IMPES is shown in Fig. \ref{fig13} \cite{12}. In this case we focus on Machines 2 and 3. The IMNEC of Machines 2 and 3 is shown in Fig. \ref{fig14}. 

\begin{figure} [H]
  \centering 
  \subfigure[]{%
  \label{fig13a}
    \includegraphics[width=0.47\textwidth]{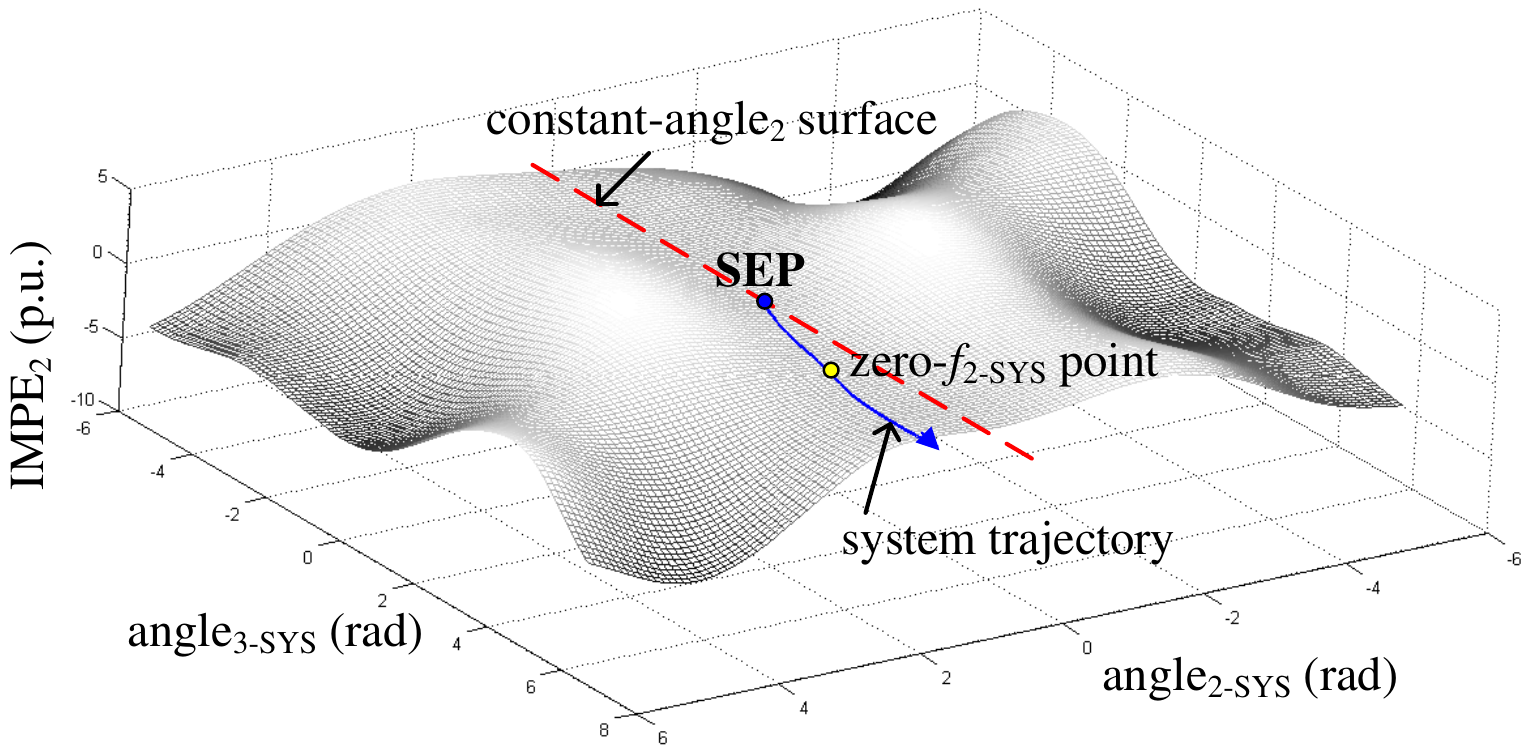}}%
\end{figure} 
\addtocounter{figure}{-1}       
\begin{figure} [H]
  \addtocounter{figure}{1}      
  \centering 
  \subfigure[]{%
    \label{fig13b}
    \includegraphics[width=0.47\textwidth]{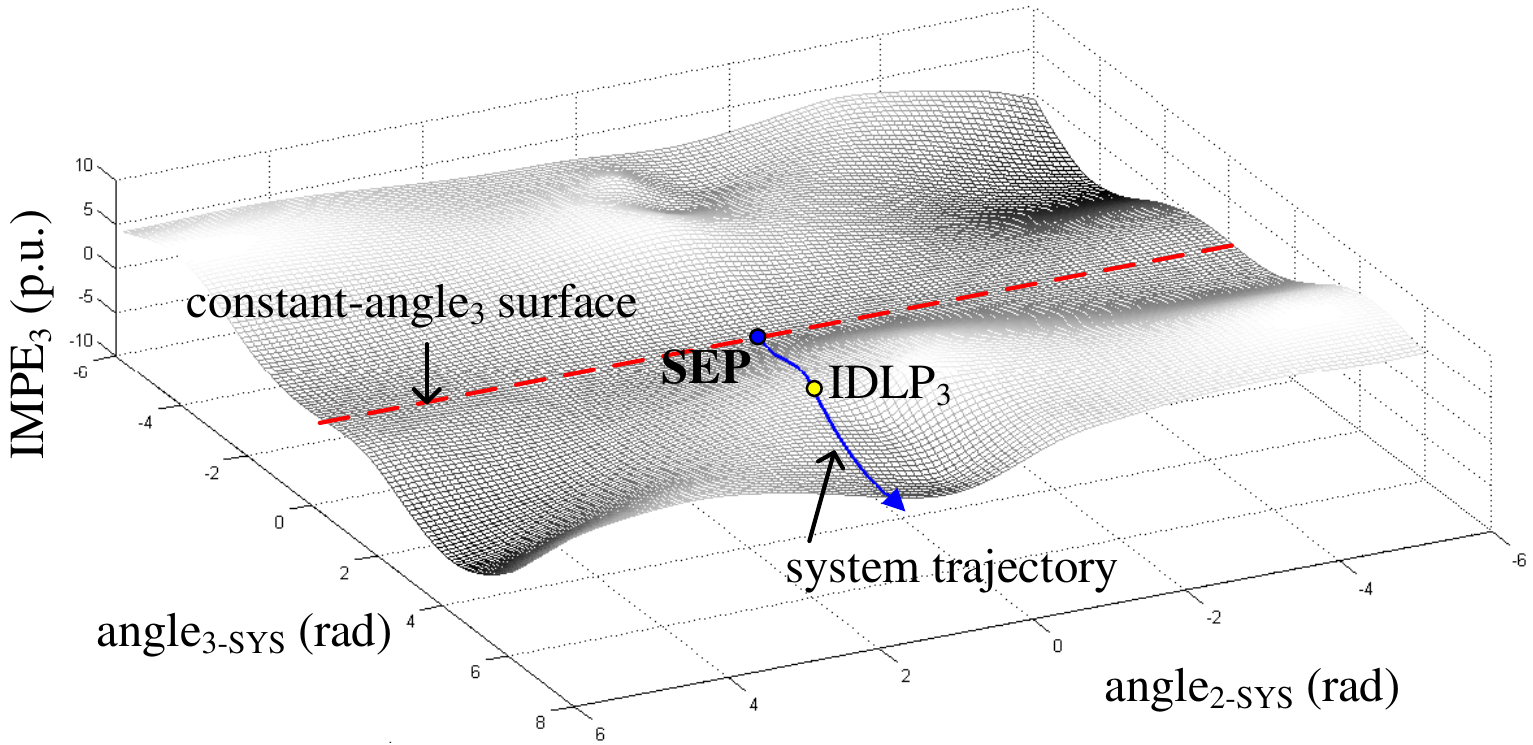}}%
  \caption{Formations of IMPES using TS-4 as the test bed.}%
  \label{fig13}
\end{figure}

\begin{figure} [H]
  \centering 
  \subfigure[]{%
  \label{fig14a}
    \includegraphics[width=0.45\textwidth]{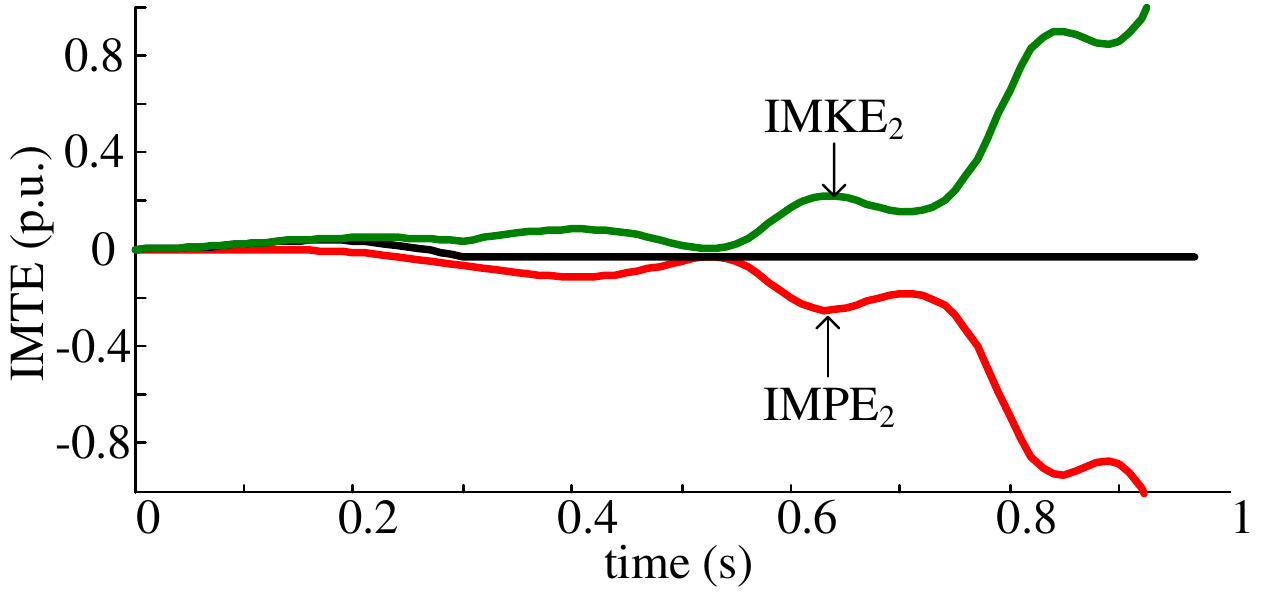}}%
\end{figure} 
\addtocounter{figure}{-1}       
\begin{figure} [H]
  \addtocounter{figure}{1}      
  \centering 
  \subfigure[]{%
    \label{fig13b}
    \includegraphics[width=0.45\textwidth]{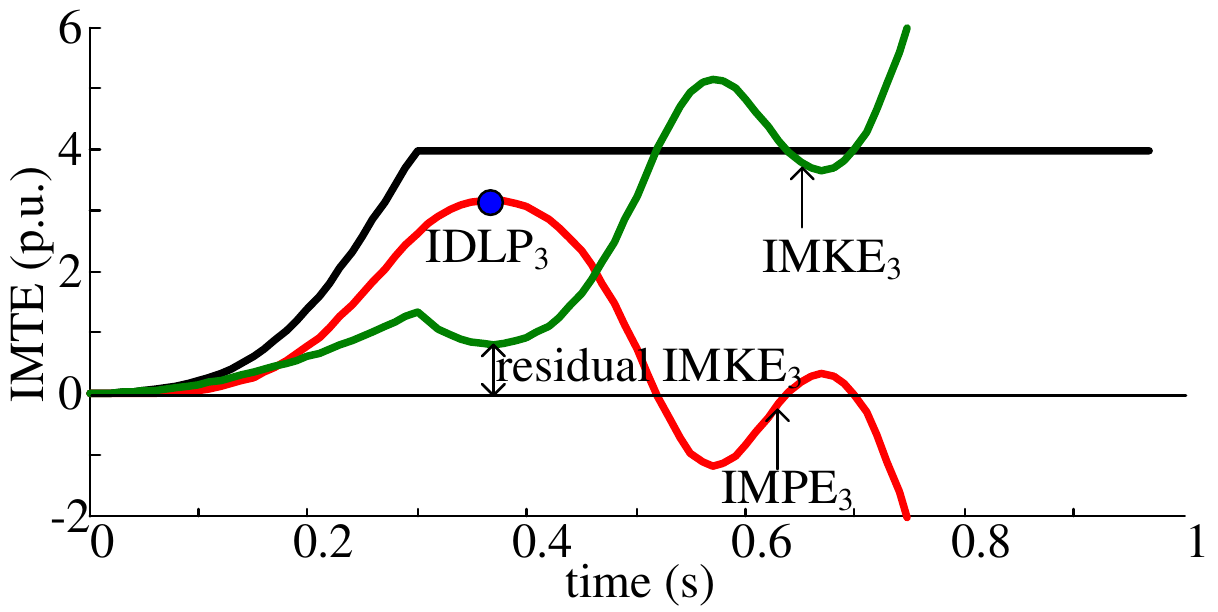}}%
  \caption{Transient energy conversion inside an individual machine. (a) Machine 2. (b) Machine 3.}%
  \label{fig14}
\end{figure}
From Fig. \ref{fig13}, the characteristics of the IMPES are summarized as below.
\\
\\ \textit{Characteristic-I}: The ``trajectory” of the ball is the system trajectory and it is defined in the $\theta_2\mbox{-}\theta_3$ angle space.
\\ \textit{Characteristic-II}: The ``altitude” of the ball is the IMPE of the machine.
\\
\par Based on the two characteristics above, because the IMPES is completely formed by the IMPE of the machine, this transient energy conversion inside the energy ball shows strict NEC characteristic.

\section{CASE STUDY} \label{section_V}
\subsection{TEST BED} \label{section_VA}
\begin{figure}[H]
  \centering
  \includegraphics[width=0.42\textwidth,center]{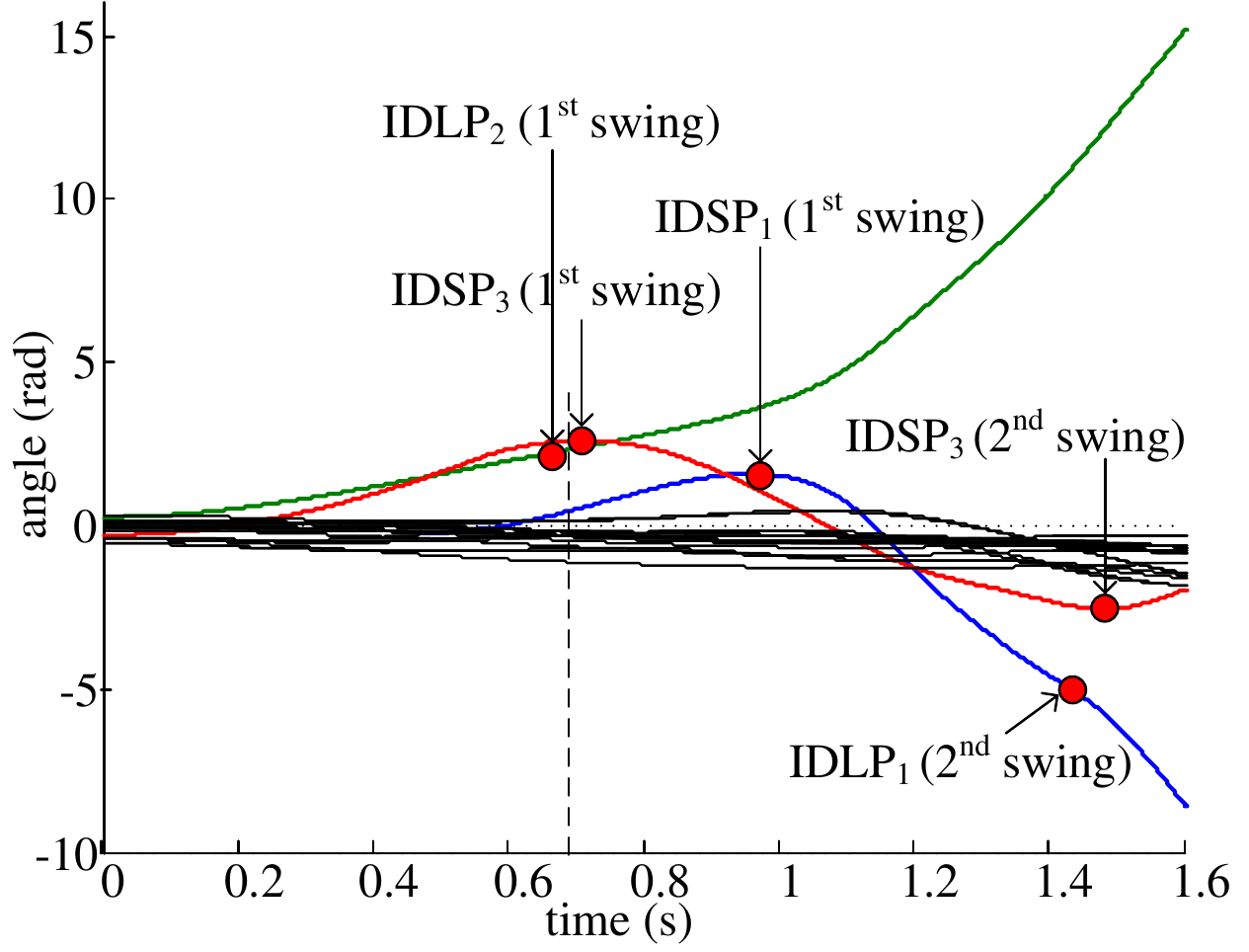}
  \caption{System trajectory [TS-2, bus-12, 0.550s].} 
  \label{fig15}  
\end{figure}

A complicated simulation case is provided to further demonstrate the two advantages of the individual-machine based TSA. The original system trajectory is shown in Fig. \ref{fig15}. Machines 1, 2 and 3 are critical machines in this case.
The application of the individual-machine in TSA is given as below.
\\
\textit{Trajectory monitoring}: Using COI-SYS as the motion reference, the system engineer monitors the variance of the IMTR of each critical machine (Machines 1, 2 and 3). Note that the system engineer monitors each critical machine in parallel.
\\ \textit{I-SYS system modeling}: The variance of the IMTR of each critical machine is modeled through its two-machine system. In this case three two-machine systems, i.e., $\text{I-SYS}_1$, $\text{I-SYS}_2$ and $\text{I-SYS}_3$ are formed.
\\ \textit{Individual-machine stability evaluation}: Following the parallel monitoring of each critical machine, the stability of each machine is evaluated independently. Based on the IMEAC, the stability of each critical machine is evaluated as below
\\
\\ \textit{$IDLP_2$ $\textit{1}^{st}$ swing occurs (0.674 s)}: Machine 2 becomes first-swing unstable. $\text{IDLP}_2$ is defined as the leading LOSP of the system.
\\ \textit{$IDSP_3$ $\textit{1}^{st}$ swing occurs (0.713 s)}: Machine 3 maintains first-swing stable.
\\ \textit{$IDSP_1$ $\textit{1}^{st}$ swing occurs (0.961 s)}: Machine 1 maintains first-swing stable.
\\ \textit{$IDLP_1$ $\textit{2}^{nd}$ swing occurs (1.415 s)}: Machine 1 becomes second-swing unstable.
\\ \textit{$IDSP_3$ $\textit{2}^{nd}$ swing occurs (1.485 s)}: Machine 3 maintains second-swing stable.
\\
\par From analysis above, because the individual-machine based TSA strictly follows the machine paradigms. the complicated characteristics of each critical machine are depicted precisely and clearly.
\\ \textit{Original system stability evaluation}: We further extend the individual-machine to the original-system level. In the TSA environment, the stability evaluation of the system is given as below
\\ The stability of the original system: At the moment that $\text{IDLP}_2$ ($\textit{1}^{st}$ swing) occurs, $\text{IDLP}_2$ is defined as leading LOSP, and thus the original system is considered to become unstable according to the unity principle.
\\ The severity of the original system: At the moment that $\text{IDLP}_1$ ($\textit{2}^{nd}$ swing) occurs, the system engineer finally confirms that Machines 2 and 1 become unstable. The severity of the entire original system is obtained as two machines becoming unstable.
\par From analysis above, the individual-machine show the two advantages in TSA. That is, (i) the stability of each machine is characterized precisely, and (ii) the trajectory of each machine is also depicted clearly at IMPP. This is because the individual-machine is fully based on the machine paradigms. 
Extending to the system level, the stability of the system is characterized in a ``machine-by-machine” manner. The instability and the severity of the original system are obtained at $\text{IDLP}_9$ and $\text{IDLP}_8$, respectively.
The key reason for the ``machine-by-machine” stability characterization is that the transient characteristics of each machine is unique and different.

\subsection{MULTI-SWING STABILITY CHARACTERIZATION} \label{section_VB}
The two advantages of the individual-machine can be fully reflected in the multi-swing stability characterization of the machine.
\par
\begin{figure}[H]
  \centering
  \includegraphics[width=0.45\textwidth,center]{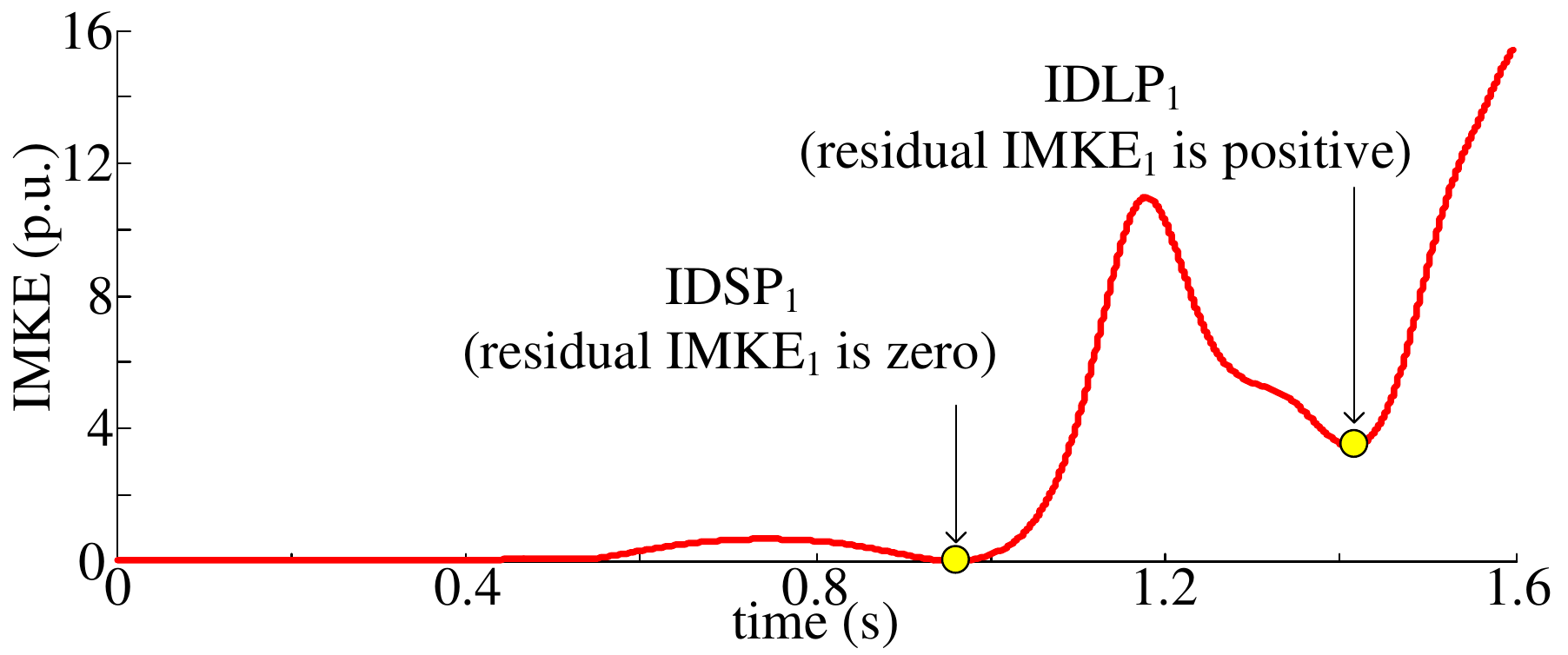}
  \caption{Variance of $\text{IMKE}_1$ along time horizon.} 
  \label{fig16}  
\end{figure}

\begin{figure}[H]
  \centering
  \includegraphics[width=0.45\textwidth,center]{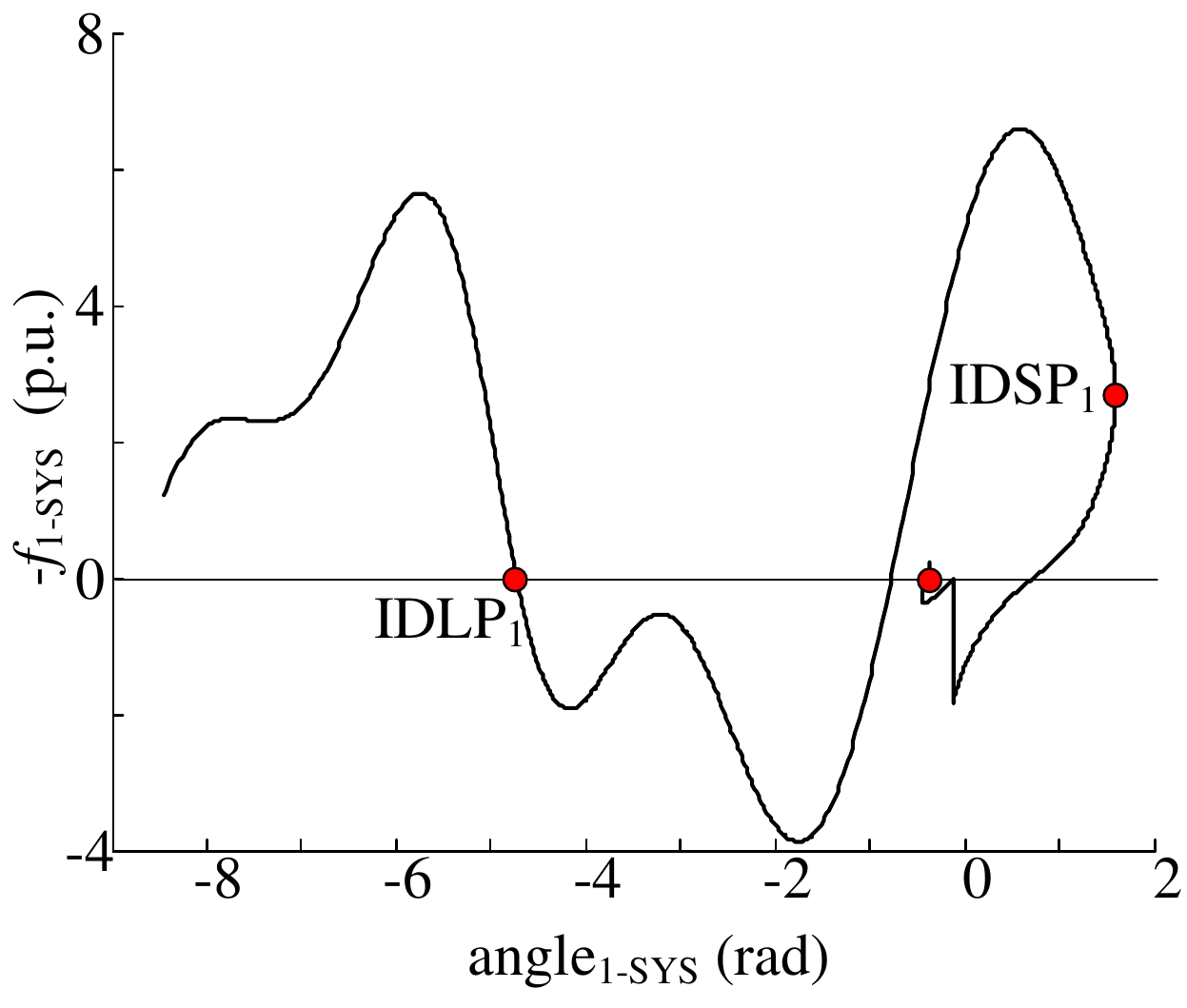}
  \caption{Kimbark curve of Machine 1.} 
  \label{fig17}  
\end{figure}

 The description of the $\text{IMTR}_1$ through $\text{IMPP}_1$ is already given in Fig. \ref{fig1}.
The individual-machine transient energy conversion inside Machine 1 is shown in Fig. \ref{fig16}. The Kimbark curve of the machine is shown in Fig. \ref{fig17}.

From Figs. \ref{fig16} and \ref{fig17}, following the IMNEC inside Machine 1, the machine is evaluated to be first-swing stable with zero residual $\text{IMKE}_1$, and it is evaluated to become second-swing unstable with positive residual $\text{IMKE}_1$.
The first swing stability and the second swing instability in the variance of the $\text{IMTR}_1$ is also depicted precisely through $\text{IDSP}_1$ and $\text{IDLP}_1$, respectively, as in Fig. \ref{fig14}.
\par From analysis above, based on the strict followings of the paradigms, the stability of each critical machine is characterized precisely, and the trajectory of each machine is depicted clearly.
The two advantages fully ensures the strict correlation between the IMTR variance of the machine and the individual-machine transient energy conversion in the original system, even though the IMTR variance of the machine is complicated.

\section{CONCLUSIONS} \label{section_VI}
In this paper the mechanisms of the individual-machine method are analyzed. Through these strict paradigm-followings, the individual-machine shows the two advantages in TSA. That is, the stability of each machine is characterized precisely, and the trajectory of each machine is depicted clearly. The two advantages can also be clearly found in the precise definitions of individual-machine based transient stability concepts.
In particular, the trajectory-depiction advantage is reflected in the definition of the critical-machine swing; Both the stability-characterization advantage and the trajectory-depiction advantage are reflected in the definition of the critical stability of the system; The stability-characterization advantage is also shown in the modeling of the IMPES. Simulation results show that the two advantages of the individual-machine based TSA can be fully reflected in the multi-swing stability characterization of the machine.
\par In the following paper, the mechanisms and ``energy superimposition” based superimposed machine will be analyzed. This may explain the inherit problems of the superimposed machine in TSA.

%

%
%
%




\end{document}